\documentclass[fleqn,usenatbib]{mnras}

\usepackage{mathptmx} 

\usepackage[T1]{fontenc}
\usepackage{ae,aecompl}

\usepackage[flushleft]{threeparttable}

\usepackage{graphicx}	
\usepackage{amsmath}	
\usepackage{amssymb}	
\usepackage{lscape}
\usepackage{siunitx}
\usepackage{booktabs,caption}
\usepackage{subfig}

\usepackage{xcolor} 

\title[Radio emission from V445~Pup]{Radio light curves and imaging of the helium nova V445 Puppis reveal seven years of synchrotron emission}

\author[Nyamai et al.]{M.~M.\ Nyamai$^{1}$\thanks{nymmir001@myuct.ac.za}, 
L.\ Chomiuk$^{2}$\thanks{chomiuk@pa.msu.edu},
V.~A.~R.~M.\ Ribeiro$^{3,4}$,
P.~A.\ Woudt$^{1}$, 
J.\ Strader$^{2}$, and
\newauthor
K.~V.\ Sokolovsky$^{2,5}$ 
\\
$^{1}$Department of Astronomy, University of Cape Town, Private Bag X3, Rondebosch 7701, South Africa\\
$^{2}$Department of Physics and Astronomy, Michigan State University, East Lansing, MI 48824, USA\\
$^{3}$Instituto de Telecomunica\c{c}\~oes, Campus Universit\'ario de Santiago, 3810-193 Aveiro, Portugal \\
$^{4}$Departamento de F\'isica, Universidade de Aveiro, Campus Universit\'ario de Santiago, 3810-193 Aveiro, Portugal \\
$^{5}$Sternberg Astronomical Institute, Moscow State University, Universitetskii~pr.~13, 119992~Moscow, Russia}

\date{Accepted XXX. Received YYY; in original form ZZZ}

\pubyear{2019}
\begin{document}
\label{firstpage}
\pagerange{\pageref{firstpage}--\pageref{lastpage}}
\maketitle

\begin{abstract}


V445 Puppis is the only helium nova observed to date; its eruption in late 2000
showed high velocities up to 8500 km s$^{-1}$, and a remarkable bipolar morphology cinched by an equatorial dust disc. Here we present multi-frequency radio observations of V445~Pup obtained with the Very Large Array (VLA) spanning 1.5--43.3 GHz, and between 2001 January and 2008 March (days $\sim$89--2700 after eruption). The radio light curve is dominated by synchrotron emission over these seven years, and shows four distinct radio flares. Resolved radio images obtained in the VLA's A configuration show that the synchrotron emission hugs the equatorial disc, and comparisons to near-IR images of the nova clearly demonstrate that it is the densest ejecta---not the fastest ejecta---that are the sites of the synchrotron emission in V445~Pup. The data are consistent with a model where the synchrotron emission is produced by a wind from the white dwarf impacting the dense equatorial disc, resulting in shocks and particle acceleration. The individual synchrotron flares may be associated with density enhancements in the equatorial disc and/or velocity variations in the wind from the white dwarf. This overall scenario is similar to a common picture of shock production in hydrogen-rich classical novae, but V445~Pup is remarkable in that these shocks persist for almost a decade, much longer than the weeks or months for which shocks are typically observed in classical novae.

\end{abstract}




\section{Introduction}
Extensive accretion of helium-rich material onto a white dwarf (WD) from a helium companion star results in increasing density and temperature on the WD surface, eventually triggering a helium shell flash (known as a helium nova; \citealt{kato1989helium}). Helium novae occur for mass accretion rates in the range $\sim 10^{-8}~\rm to~10^{-7}~ M_\odot$ yr$^{-1}$ 
 \citep{woosley2011sub, Piersanti2013}. The thermonuclear runaway ejects the outer accreted layers analogous to hydrogen-rich classical novae \citep[see e.g.,][]{bode2008classical, woudt2014stella}, although more mass is ejected in helium novae \citep{kato1989helium, hillman2016growing}. However, when the helium accretion rate is lower ($\leq~ 10^{-8}~\rm M_\odot$ yr$^{-1}$), it is possible for 
 a helium ignition on the surface of the WD to trigger an inward shockwave which leads to an explosion at the core of the WD hence a double detonation \citep{nomoto1982, Shen2009}. If the detonation occurs on a carbon-oxygen WD, it could produce a Type Ia supernova  (SN Ia; \citealt{moll2013multi, ruiter2014effect, Piersanti2014}). The study of helium accretion is therefore important for validating or ruling out helium-donor and WD binary systems as progenitors of SNe Ia. 
 

In this paper, we present radio observations of the only spectroscopically confirmed helium nova, V445~Puppis.
The nova was discovered in eruption in late November 2000 at an optical maximum of $V\approx 8.6$ mag \citep{Kato2000}, substantially brighter than its $V\approx14.5$ pre-eruption magnitude \citep{ashok2003}. The day of eruption for V445 Pup is unknown and only constrained to be between 2000 September 26 and 2000 November 28 \citep{Ashok2001, Kato2000, woudt2009}. We adopt a $t_0$ from \citealt{woudt2009} of 2000 November 02 (MJD = 51850) as the time of eruption. The nova was unusually rich in carbon and showed helium emission lines, but lacked hydrogen lines (which are prominent in the spectra of classical novae; \citealt{ashok2003, iijima2008, woudt2009}).

Within one month of the eruption, dust was detected from the nova using infrared spectroscopy \citep{lynch2004}. By $\sim$8 months after the nova eruption, the optical brightness had dropped below pre-eruption levels \citep{ashok2003, woudt2009}, attributed to a strong dust formation episode. Six years following the eruption, the optical light had still not returned to pre-eruption levels \citep{woudt2009}, and the dust mass was determined to be $> 10^{-5}~\rm M_\odot$ \citep{shimamoto2017infrared}. 
 
V445~Pup was spatially resolved using near-infrared adaptive optics imaging with the Very Large Telescope (VLT), revealing an expanding bipolar shell confined by an equatorial dust disc and with polar knots on both ends of the bipolar ejecta (\citealt{woudt2009}; see their Figure 2). \citet{woudt2009} determined a distance of $8.2~\pm~0.5~{\rm~kpc}$ to the nova using expansion parallax techniques.

Presented in this paper are multifrequency observations of the nova obtained in the years following eruption using the Very Large Array (VLA) radio telescope. The radio data reveal an unprecedented, near decade-long, synchrotron-powered radio light curve. In \S 2, we discuss the radio observations and our data analysis procedure. In \S 3, we present the multi-frequency radio light curve, radio spectral evolution, and spatially-resolved radio imaging. In \S 4, we present the results, and in \S 5 we highlight our conclusions.

\section{Observations and data analysis}
\label{sec:Methods} 
 Radio observations of V445 Pup were obtained with a variety of VLA programs, most of which were led by M.\ Rupen. The data were retrieved from the VLA archive (see Table~\ref{tab:Table_obs} for a log of the observations). 
 The first radio observations of V445~Pup were taken on 2001 January 18 and 30 ($t-t_0$ = 77 and 89 days). On day 77, the nova was not detected at 8.4 GHz \citep{Rupen2001a}, and there was a marginal detection on day 89. After day 89, observations were paused until 2001 September 9 (day 312 day), when the nova was strongly detected at 8.4 GHz \citep{Rupen2001a}. V445~Pup was subsequently observed with the VLA from September 2001 to March 2008 (between day $312-2704$), resulting in a detailed radio light curve spanning almost a decade following the eruption (Figure~\ref{fig:full_lc}).

 Observations were obtained in continuum mode, at different frequency bands with two 50 MHz-wide frequency channels.
The observations were conducted at 
L (1.46 GHz), C (4.86 GHz), X (8.46 GHz), U (14.94 GHz), K (22.46 GHz) and Q (43.34 GHz) bands, and are used to trace the nova emission throughout all of the VLA's configurations. The VLA A configuration provides the highest resolution (synthesized beam FWHM of $2\arcsec$
at 1.5 GHz and $0.3\arcsec$ at 8.5 GHz), and has the potential to provide resolved images (see \S~\ref{sec:image}). At each frequency, observations of the target are obtained together with observations of gain calibrators (see Table~\ref{tab:Table_phase}). The most common gain calibrator used was 0804-278, for which we took the ICRS coordinates RA = 08h04m51.451s and Dec = $-27^{\circ}49^{\prime}11.32^{\prime\prime}$. We note that a slightly different RA position was used by National Radio Astronomy Observatory (NRAO) for this calibrator in 2001--2002 (RA = 08h04m51.440s); we, therefore, shifted the data so that all observations take RA = 08h04m51.451s (this is particularly relevant for our imaging results; \S \ref{sec:image}). Flux density calibrators 0137$+$331 (3C48), 0542$+$498 (3C147) or 1331$+$305 (3C286) were observed to set the absolute flux density scale. Some observations were obtained without a flux density calibrator; in most cases, we do not include these observations in this work. For the ones included, we used a flux calibrator of an adjacent epoch to set the flux density of the secondary calibrator.

\begin{table*}
    \centering
    \caption{Log of VLA observations of V445~Pup.}
    \label{tab:Table_obs}
    \begin{threeparttable}
\begin{tabular}{lccccccccr}
\hline
Observation & $t$ & $t-t_0$ & Configuration & \multicolumn{6}{c}{Observation time on target (min)}\\
Date & (MJD) & (Days) & &1.43 GHz & 4.86 GHz & 8.46 GHz& 14.94 GHz & 22.46 GHz & 43.34 GHz \\ 
\hline 
2001 Jan 18 & 51927 & 77 & A &... & ...& 3.7 & ...&... &...  \\
2001 Jan 30 & 51939 & 89 & A &... & ...& 6.4 & ...&... &...  \\
2001 Sept 09 & 52162 & 312 & C &... & ...& 6.7 & ...&... &...  \\
2001 Sept 11 & 52164 & 314 & C & 15.6 & 3.7&... & 5.2 &4.6 &4.4 \\
2001 Sept 12 & 52165 & 315 & C &10.9 & 3.9 & 3.7 & 5.2& 4.4 &... \\
2001 Sept 14 & 52167 & 317 & C &... & 4.1&... & ... &... & ...\\
2001 Sept 15 & 52168 &318 & C\&D&... & 1.9& 3.1& ... &... &... \\
2001 Sept 16 & 52169 &319 & C\&D & 5.2 & 6.9& 6.9& 4.9 & ... &... \\
2001 Sept 17 & 52170 &320 & C\&D& 10.4 &11.2& ...& 11.9&... & ...\\
2001 Sept 20 & 52173 & 323& C\&D&3.7 &4.2&4.2& 4.2 &4.2 &... \\
2001 Sept 25 & 52178 & 328 & C\&D &... & 11.7 & 12.7&14.8 &... &... \\
2001 Sept 26 & 52179 & 329 & C\&D &... & 3.2& 4.7& 6.2 &6.2 & ...\\
\hline 
\end{tabular}
\end{threeparttable}
\begin{tablenotes}
\item `...' indicates no observations for this epoch at that frequency. Here, $t_0$ is taken as 2000 November 02 (MJD = 51850). This table is continued in Appendix \ref{sec:observations_and_fluxes} Table~\ref{tab:continued_VLA_obs}. 
\end{tablenotes}
\end{table*}

\begin{table*}
    \centering
    \caption{Log of gain calibrators used at different configurations and observing frequencies.}
    \label{tab:Table_phase}
    \begin{threeparttable}
\begin{tabular}{lcccccccr}
\hline
Observation &   &\multicolumn{6}{c}{Gain calibrators}\\
Date range & Configuration &1.43 GHz & 4.86 GHz & 8.46 GHz& 14.94 GHz & 22.46 GHz & 43.34 GHz\\ 
\hline 
2001 Jan 18--2001 Jan 30& A& ... &... & 0804-278&... &... &...& \\
2001 Sep 09--2001 Sep 14 & C & 0738-304& 0804-278 & 0804-278& 0804-278 & 0804-278 & 0804-278 \\
2001 Sep 15--2001 Oct 10 & C\&D & 0735-175 & 0804-278 & 0804-278& 0804-278 & 0804-278 & 0804-278 \\
    &   & 0806-268 &  0738-304 & & & & \\
        &   &  0706-231&  & & & & \\
2001 Oct 12-- 2002 Jan 13 & D & 0706-231 & 0738-304,  & 0804-278& 0804-278 & 0804-278 & 0804-278 \\
    &   & 0806-268 &  0804-278 & & & & \\
2002 Feb 07-- 2002 Jun 07 & A, A\&B & 0804-278 &0804-278 & 0804-278& 0804-278& 0804-278&... \\
2002 Jun 17-- 2002 Oct 31 & B & 0804-278 & 0804-278 & 0804-278&... &... & 0804-278\\
2002 Dec 09-- 2003 Jan 29 & C, C\&D &... &0804-278 & 0804-278&... & ...&... \\
2003 Feb 06-- 2003 Apr 18 & D &... & 0804-278 & 0804-278&... & ...& ...\\
2003 May 30-- 2003 Oct 08 & A, A\&B &0735-175 & 0804-278 & 0804-278&... &... &... \\
        &   &  0804-278 &  & & & & \\
2003 Oct 21-- 2004 Mar 02 & B, B\&C &0804-278 &0804-278 & 0804-278& 0804-278& 0804-278 & 0804-278\\
2003 Mar 10-- 2004 June 12 & C, C\&D &0738-304 & 0804-278 & 0804-278&... & 0804-278 &... \\
        &   & 0706-231 &  & & & & \\
2004 Jun 25-- 2004 Aug 21 & D &... &0804-278 & 0804-278& 0804-278&... & ...\\
2004 Sep 09-- 2005 Feb 03 & A, A\&B &... & 0804-278 & 0804-278&0804-278 &... &... \\
2005 Feb 26-- 2005 Jul 02 & B, B\&C &0804-278 & 0804-278& 0804-278& ...& ...&... \\
2005 Jul 09-- 2005 Oct 31 & C, C\&D &... &0804-278 & 0804-278& ...&0804-278 &... \\
2005 Nov 08-- 2006 Jan 22 & D &... & 0738-304& 0804-278& ...&... & ...\\
2006 Mar 15-- 2006 May 10 & A &... &... & 0804-278&... &... &... \\
2006 Jun 25-- 2006 Sep 11 & B &... &0804-278 & 0804-278&... &... & ...\\
2007 Sep 30--2008 Jan 17 & A\&B, B &... &... & 0804-278&... &... &... \\
2007 Jan 23, 2008 Mar 28& C\&D, D & ...&... & 0804-278&... & ...&... \\
\hline 
\end{tabular}
\end{threeparttable}
\begin{tablenotes}
\item `...' indicates no observations at that date range and frequency. 
\end{tablenotes}
\end{table*}

All data were processed using the Common Astronomy Software Applications (CASA; \citealt{McMullin2007}). The \texttt{AOflagger} algorithm \citep{offringa2012} was used to flag data corrupted by radio frequency interference. Standard calibration procedures were applied to each observation, and calibration solutions were applied to the target data before imaging. Starting in 2006, the VLA antennas were gradually upgraded to Jansky VLA capabilities, and we therefore found baseline-based calibration solutions using the flux calibrator.
The CASA task \texttt{clean} was used for imaging, utilising Briggs weighting with a robust value of 1. Self-calibration was not performed.

To measure the flux densities of V445~Pup, the CASA task \texttt{imfit} was used to fit a Gaussian to the nova in each resulting image. In most measurements the width of the Gaussian is allowed to vary, and the integrated flux density was recorded.
However, in cases of low signal-to-noise, the size of the Gaussian was fixed to the size of the synthesized beam. The uncertainty in flux density includes the error from the Gaussian fit added in quadrature with estimates of the uncertainty on the absolute flux calibration (5$\%$ of the flux for 1--10 GHz and 10$\%$ of the flux for frequencies greater than 10 GHz). In epochs where there is a non-detection, the task \texttt{imstat} was used to determine the noise of the image. An upper limit on the flux density is determined as three times the image rms plus the value of the pixel at the target location. 

The flux densities of V445~Pup for each frequency band are given in Table~\ref{tab:Table_fluxes}; the first ten epochs are presented in this manuscript and others are available in the electronic materials.

\section{Results}
\subsection{Radio light curves}

\begin{figure*}
	\includegraphics[width=0.98\textwidth]{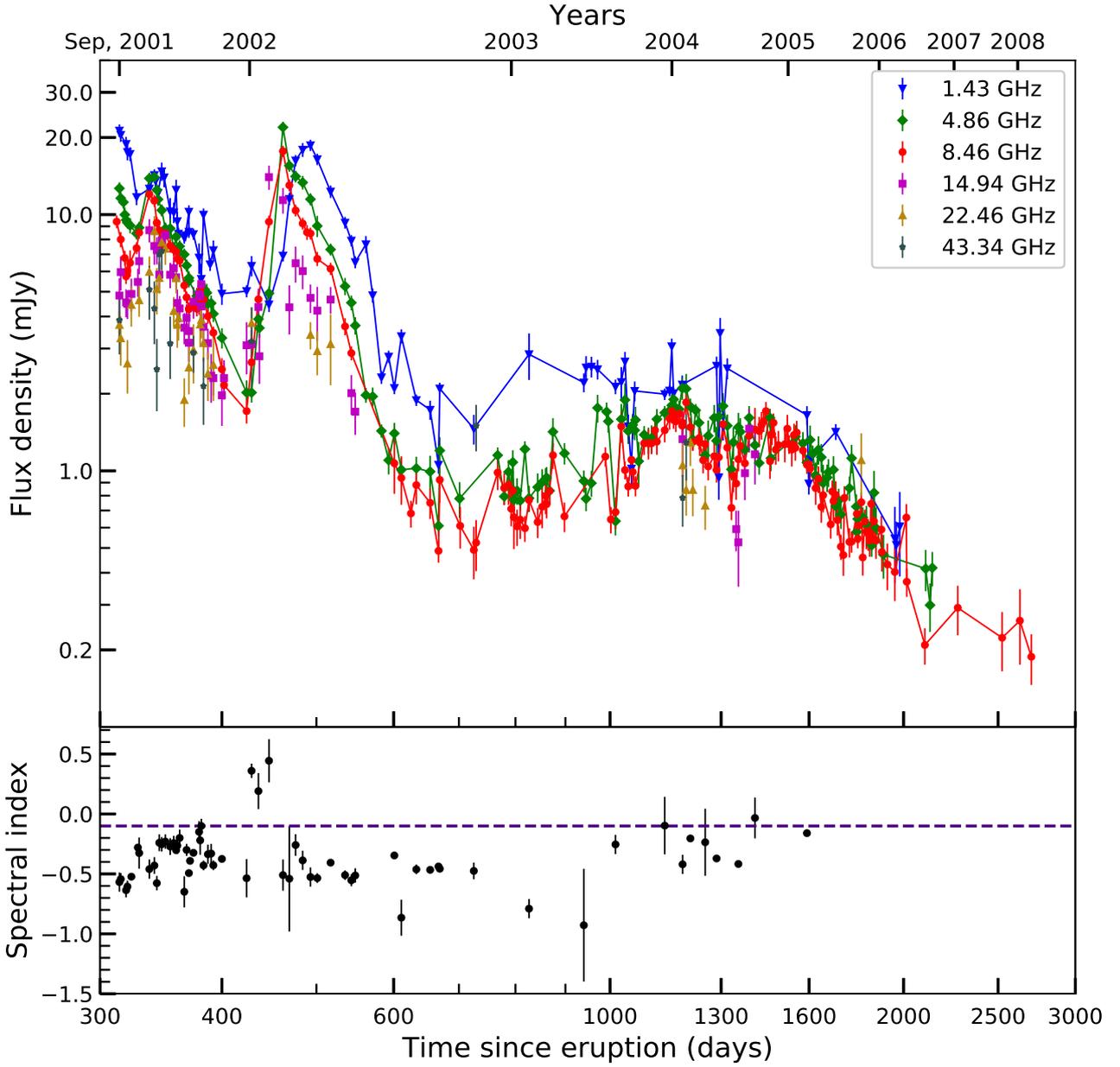}
    \caption{Top: Observed flux densities of V445~Pup spanning from day 300 to day 2700 after the nova eruption. We take 2000 November 02 as the date of the eruption ($t_0$). Bottom: spectral indices obtained by fitting a single power-law to the data. The dashed line in the lower panel represents $\alpha = -0.1$, the theoretically expected index of optically thin free-free emission.}
    \label{fig:full_lc}
\end{figure*}

Multi-frequency observations, spanning 1.5--43.3 GHz, reveal an unusual radio light curve for the eruption of V445~Pup between 2001--2008 presented in Figure~\ref{fig:full_lc}. The radio light curve shows four distinct flares throughout its evolution.

\begin{figure*}
	\includegraphics{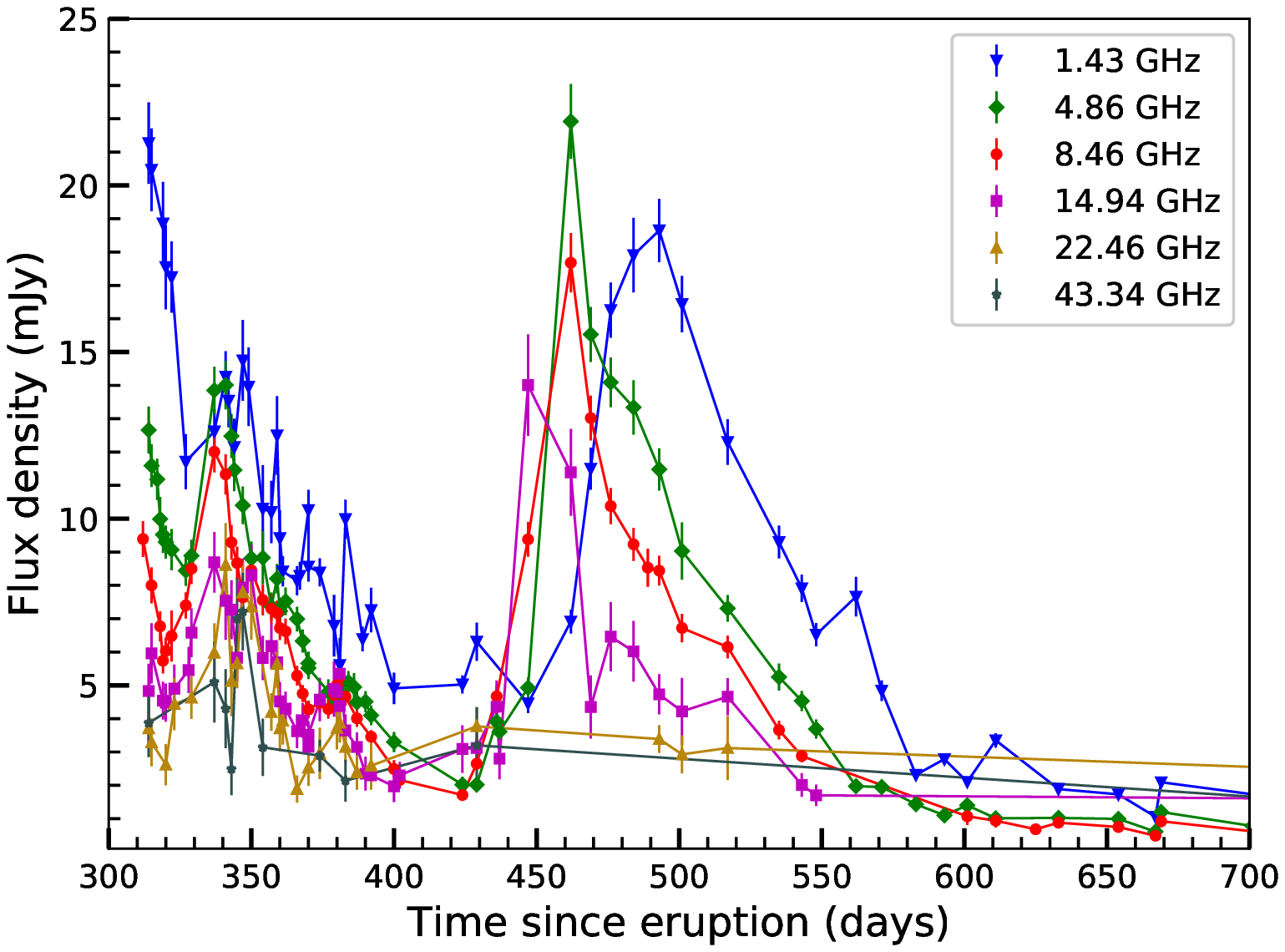}
    \caption{Observed flux densities of V445~Pup spanning from day 300 to day 700 after the nova eruption. Data are the same as those plotted in Figure \ref{fig:full_lc}, but zoomed in to show detail of the radio flares indicated with vertical dashed lines.}
    \label{fig:zoom_lc}
\end{figure*}

In the earliest radio observation following the eruption of V445~Pup, around day 77, no radio emission was detected, with a 3$\sigma$ upper limit of 0.34 mJy at 8.4 GHz. However, on day 89, a 4$\sigma$ detection of 0.35 mJy was obtained at 8.4 GHz. Unfortunately, following the day 89 detection there were no radio observations of the nova until 312 days after eruption. As a result of very limited coverage during the first 300 days after eruption, these early points are not shown in Figure~\ref{fig:full_lc}. On day 312, it became clear that V445~Pup had brightened at radio frequencies since day 89 by a factor of $\sim 26$ to 9.4 mJy at 8.4 GHz. Over the next $\sim$3 months (see Figure~\ref{fig:zoom_lc}), the light curve shows flux densities declining from a radio maximum, which presumably occurred before day 312.

Other flares are superimposed on the decline, like the steep peak on day 337 (at 4.9/8.5/14.9 GHz; Figure~\ref{fig:zoom_lc}). Another flare peaked between days 450 to 460 (at 4.86/8.46/14.94 GHz bands). We note that, for the lowest frequency (1.43 GHz) light curve, these peaks occur later, at days 341 and 493 (Figure~\ref{fig:zoom_lc}).

The radio peaks observed between day 300--700, indicated by vertical dashed lines in Figure~\ref{fig:zoom_lc},
have sharp, rapid rises and falls in flux density, compared to the last broad peak which occurs between days 700 and 3000 (Figure~\ref{fig:full_lc}).
Furthermore, the flux density at the peak of the last flare is a factor of five fainter than the early-time peaks.

\begin{table*}
    \centering
    \caption{Flux densities and spectral indices of V445~Pup}
    \label{tab:Table_fluxes}
\begin{tabular}{lcccccccr}
\hline
$t$ & $t-t_0$ &\multicolumn{6}{c}{Radio flux densities (mJy)}& $\alpha$\\
 (MJD) & (Days) &1.43 GHz & 4.86 GHz & 8.46 GHz& 14.94 GHz & 22.46 GHz & 43.34 GHz&\\ 
\hline 
51927 & 77 & ... & ... & $<$~0.34  & ...&... &...& ... \\
51939 & 89 & ... & ... & 0.354 $\pm$ 0.064 & ...&... &...& ... \\
52162 & 312 & ... & ... & 9.39 $\pm$ 0.54 & ...&... &...& ... \\
52164 & 314 &21.27 $\pm$ 1.22 &12.66 $\pm$ 0.71 & ...  &4.83 $\pm$ 0.82 & 3.71 $\pm$ 0.50&3.87 $\pm$ 1.03& $-0.57~\pm~ 0.08$\\
52165 & 315 &20.47 $\pm$ 1.24 & 11.59 $\pm $ 0.64  & 8.00 $\pm$ 0.54  & 5.96 $\pm$ 0.90 & 3.29 $\pm$ 0.72&...& $-0.54 ~\pm~ 0.05$ \\
52167 & 317 & ...& 11.18 $\pm$ 0.62 &... & ... &... &... &...\\
52168 & 318  &... &9.99 $\pm$ 0.66 & 6.77 $\pm$ 0.45& ... &... & ...&...\\
52169 & 319 &18.86 $\pm$ 1.24 & 9.52 $\pm$ 0.55& 5.74 $\pm$ 0.38 & 4.54 $\pm$ 0.57 &... &... & $-0.64 ~\pm~ 0.06$\\
52170 & 320 &17.55 $\pm$ 1.26& 9.30 $\pm$ 0.50 &6.03 $\pm$ 0.41 &4.48 $\pm$ 0.57 &2.62 $\pm$ 0.62  &... &$-0.61 ~\pm~ 0.05$\\
52173 & 323&17.25 $\pm$ 1.07 & 9.07 $\pm$ 0.62 & 6.48 $\pm$ 0.76 &4.90 $\pm$ 0.72&4.44 $\pm$ 0.79 &... &$-0.52~\pm~ 0.02$\\
52178 & 328 & 11.71 $\pm$ 0.82 & 8.44 $\pm$ 0.45& 7.40 $\pm$ 0.39 &5.46 $\pm$ 0.69&... &... & $-0.28 ~\pm~ 0.03$\\
52179 & 329 &... & 8.89 $\pm$ 0.48 & 8.50 $\pm$ 0.46&6.58 $\pm$ 0.73 &4.63 $\pm$ 0.64&...& $-0.33 ~\pm~ 0.13$\\
\hline 
\end{tabular}
\begin{tablenotes}
\item `...' indicates no measurements for flux density for the epoch at that frequency. This table is continued in Appendix \ref{sec:observations_and_fluxes} Table~\ref{tab:continued_VLA_fluxes}. 
\end{tablenotes}
\end{table*}

The last recorded flux density of the nova with the VLA was 0.19 mJy at 8.46 GHz on 2008 March 29 (day 2704). On 2009 October 29 and December 29 (days 3283 and 3325 respectively), V445~Pup was observed using the Giant Metrewave Radio Telescope (GMRT) at 1.28 GHz and 1.40 GHz \citep{Kantharia2012}. Radio emission was present in the first GMRT observation, with a flux density $\sim0.3\pm0.08$ mJy. The second epoch yields a non-detection with a 3$\sigma$ upper limit of 60 $\si\micro$Jy \citep{Kantharia2012}. 

\subsection{Radio spectral evolution}
\begin{figure*}
	\includegraphics[width=1.0\textwidth] {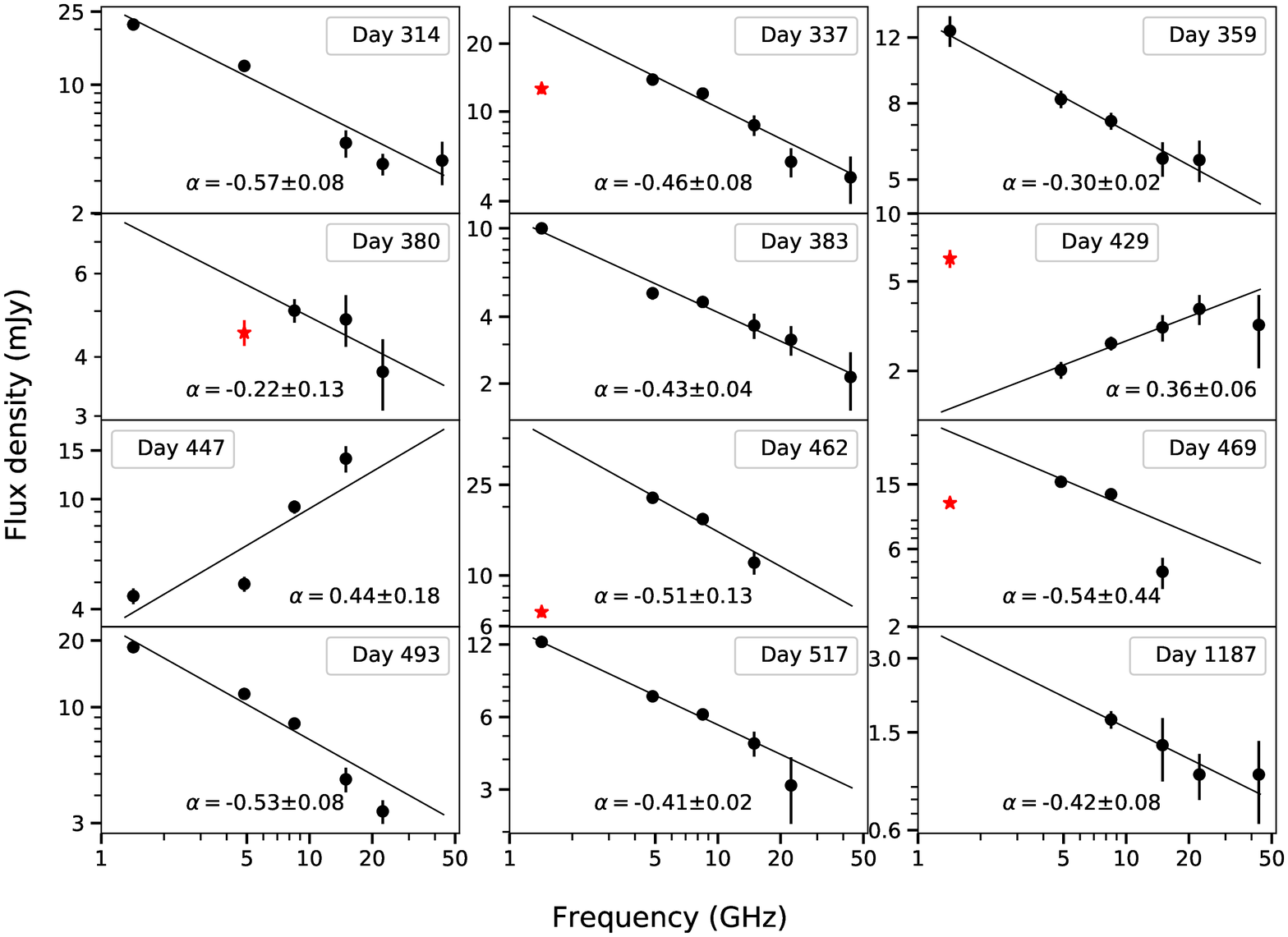}
  \caption{Selected VLA radio spectra of of V445~Pup, with linear fits overplotted. The time after eruption (since 2000 November 2) and  the value of the spectral index is given for each spectrum. The data points marked with red stars represent flux densities in L-band (1.4 GHz) or C-band (4.9 GHz) which are relatively flat and thus not included in the fit.}
  \label{fig:spectra}
\end{figure*}

We determine the spectral index for each observation ($\alpha$, defined as $S_\nu ~\propto~\nu^\alpha$, where $S_\nu$ is the flux density and $\nu$ is the observing frequency), from the slope of the line fit to the data in log-log scale. The python function \texttt{curve-fit} in the \texttt{scipy} package is used to perform the least-squares fit. Detections separated by less than a day are combined into one spectral index fit. Selected spectra are plotted in Figure~\ref{fig:spectra}, while the lower panel of Figure~\ref{fig:full_lc} shows how the spectral index varies with time.

During the first decay of the radio flux density (days $\sim$312--400), the spectrum has a form that rises steeply toward low frequencies and is well fit with a single power law. For example, on day 314, $\alpha= -0.6$ (top left panel of Figure~\ref{fig:spectra}), an indication of optically-thin synchrotron emission (see \S \ref{sec:synch} for more discussion). Sometimes the spectrum appears to flatten toward low frequency (such as on day 337) and afterwards the spectrum switches back to being well fit with a single power law (e.g., day 383) 

On days 429 and 447, the spectrum becomes inverted and rises towards higher frequencies, with $\alpha \approx 0.4$ (Figure~\ref{fig:spectra}), an indication of optically-thick emission. On days 462 and 469, the radio spectrum is transitioning back to an optically-thin state, exhibiting a combination of inverted/flat spectral index at lower frequencies and steep spectral index at higher frequencies. By day 493, the radio spectrum has returned to optically thin, hovering around $\alpha \approx -0.5$. We discuss likely causes of these changes in the radio spectrum in \S \ref{sec:lc_interp}.


\begin{figure*}
    \centering
	\subfloat{\includegraphics[width=0.49\textwidth]{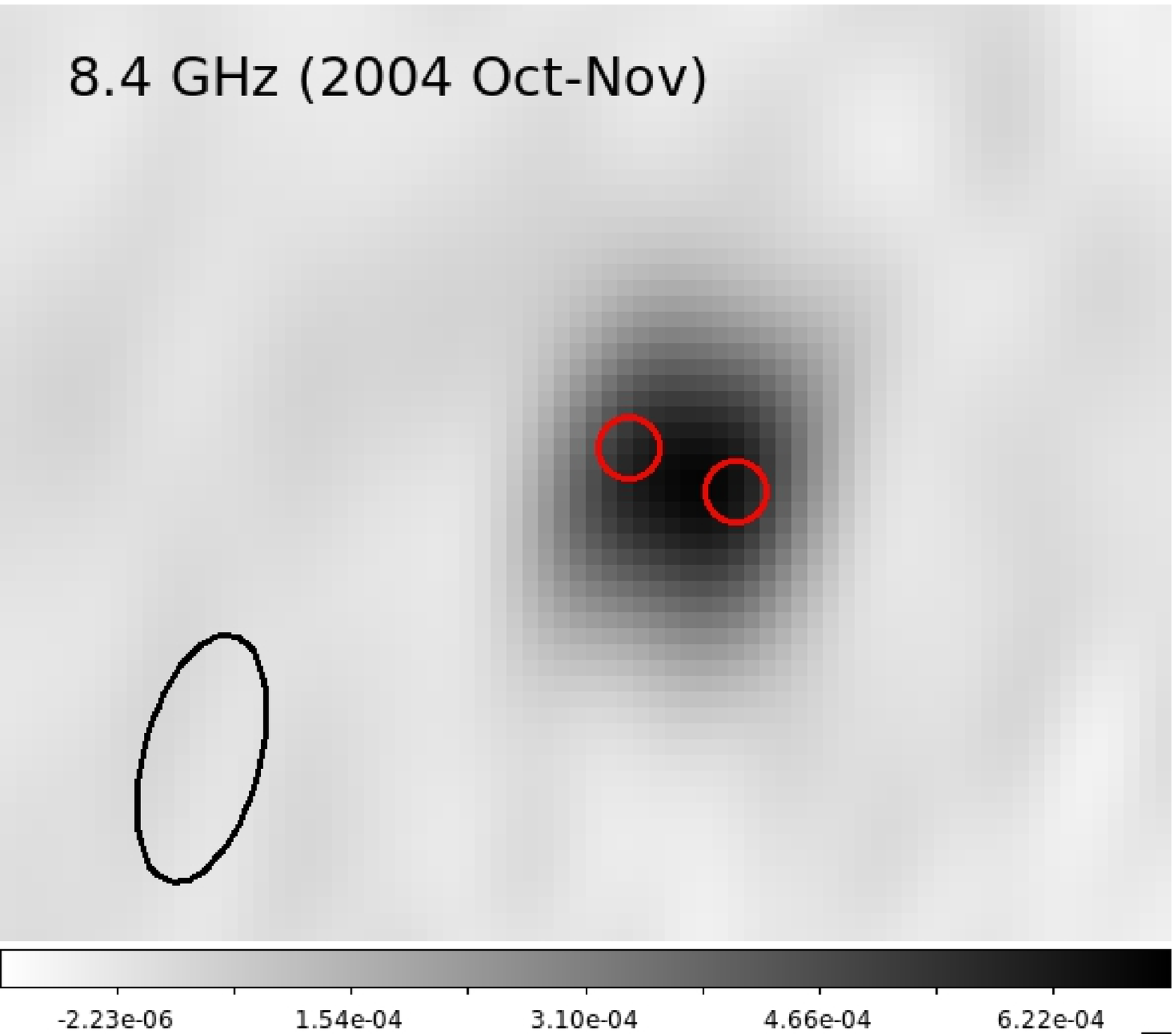}}
   \hfill
  \subfloat{\includegraphics[width=0.49\textwidth]{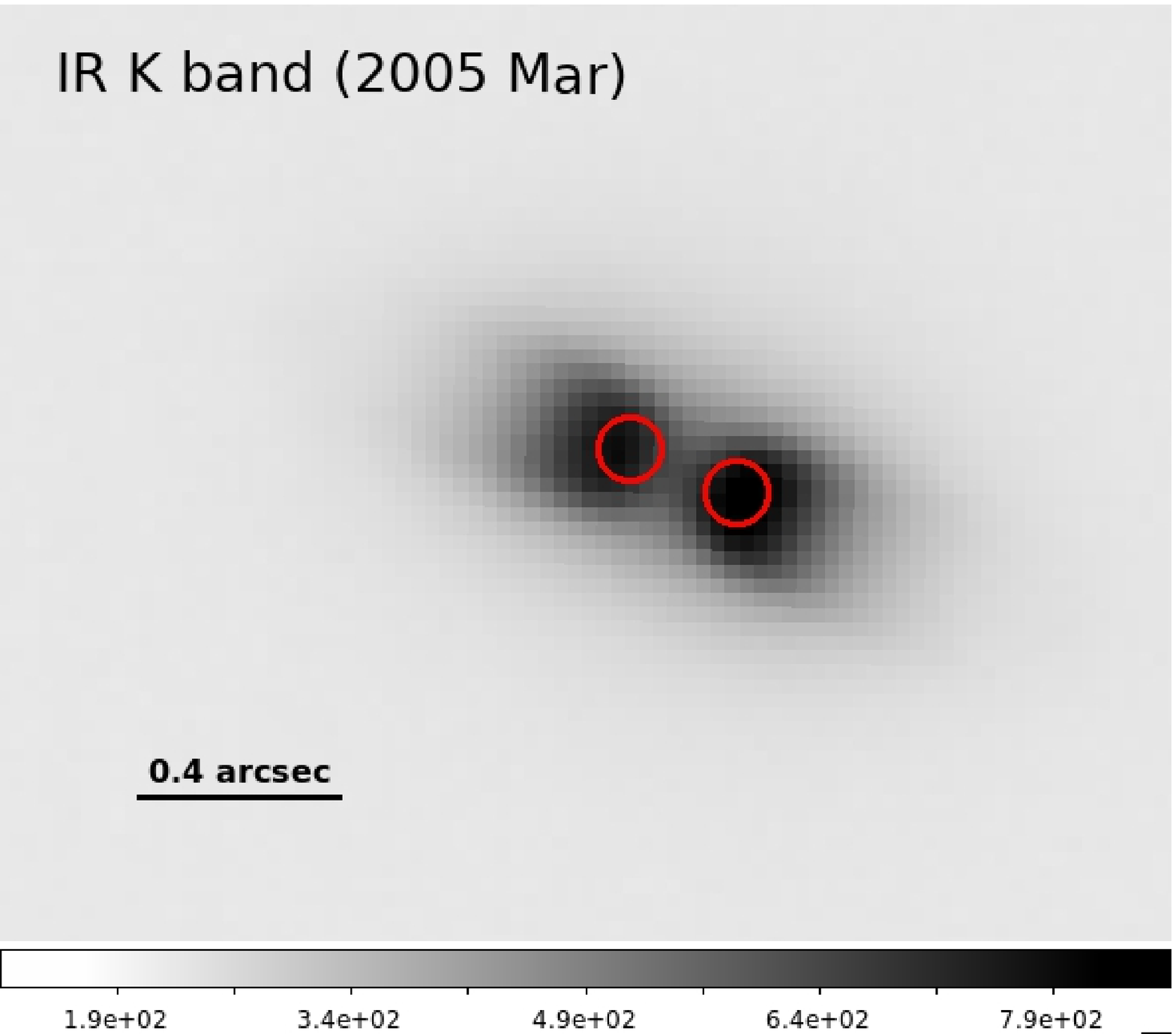}}
    \caption{High-resolution images of V445~Pup observed at radio (left) and near-infrared (right) wavelengths. The red circles are the positions of the two Gaussian components fit to the 2004 Oct--Nov A configuration data in the \textit{uv} plane, and have diameters of $0.12^{\prime\prime}$. The left panel shows our 8.4 GHz image representing the average of data  obtained between 2004 Sep 23 and 2004 Nov 29 (days 1402--1488). The ellipse in the bottom left corner shows the FWHM extent of the synthesized beam ($0.48^{\prime\prime} \times 0.23^{\prime\prime}$, PA= $165^{\circ}$).  The right panel shows a near-IR K-band image obtained on 2005 Mar 26 (day 1605), with the same positions of the radio counterparts marked as red circles. In these images, north is up and east is to the left; and the fields-of-view are matched to be $2.4^{\prime\prime} \times 1.7^{\prime\prime}$. The grey scale is linear in both cases.}
    \label{fig:hires}
\end{figure*}

\subsection{Radio Images}
\label{sec:image}

In its most extended A configuration, the VLA achieves angular resolution sufficient to constrain the morphology of radio emission from V445~Pup. 
The dates of these high-resolution observations are listed in Table~\ref{tab:aconf}, along with a brief description of the resulting images and the central position of the emission region (more details below).
We fit the morphology of V445~Pup in the \textit{uv}-plane, using the \texttt{UVFIT} task in \texttt{AIPS} \citep{2003ASSL..285..109G}  and \texttt{Difmap} \citep{1994BAAS...26..987S,1997ASPC..125...77S}.

\begin{table*}
    \centering
    \caption{High-Resolution A-configuration observations of V445~Pup.}
    \label{tab:aconf}
\begin{tabular}{lcccc}
\hline
Observation & $t-t_0$ & Description & RA (J2000.0) & Dec (J2000.0) \\
Date Range & (Days) & & (07h37m56.XXXs) &  ($-25^{\circ}56^{\prime}58^{\prime\prime}.$XX)\\ 
\hline 
2001 Jan 18--Jan 30 & 77--89 & Marginally detected & -- & --\\
2002 Feb 7--May 15 & 462--559 & Unresolved, FWHM $< 0.1^{\prime\prime}$ & $.885\pm.001$ & $.86\pm.01$\\ 
2003 May 17--Sep 19 & 926--1051 & Marginally resolved, $0.3^{\prime\prime} \times < 0.1^{\prime\prime}$ & $.880\pm.001$ & $.85\pm.02$  \\
2004 Sep 23--2005 Jan 6 & 1408--1526 & Two components separated by $0.2^{\prime\prime}$ & NE: $.890\pm.002$ & $.80\pm.03$ \\
 & & & SW: $.875\pm.002$ & $.88\pm.03$ \\
%
2006 Feb 9--May 10 & 1925--2015 & One component, $0.3^{\prime\prime} \times < 0.2^{\prime\prime}$ & $.883\pm.002$ & $.85\pm.04$ \\
2007 Jul 15--Jul 28 & 2446--2459 & Marginally detected & -- & -- \\
\hline 
\end{tabular}
\end{table*}

The VLA was in A configuration during the first observations of V445~Pup in 2001 Jan. They yielded a non-detection (Jan 17, $t-t_0~=~77$ days) and a marginal detection (Jan 30, $t-t_0~=~89$ days). The S/N was not sufficient to constrain the morphology of V445~Pup at this time.

During the 2002 A configuration (462--559 days after eruption), V445~Pup was radio bright. At 8.4 GHz, it is unresolved; model fitting with an elliptical Gaussian in the \textit{uv} plane implies FWHM $< 0.1^{\prime\prime}$.  

During the 2003 A configuration (926--1051 days after eruption), V445~Pup is marginally resolved. We model the 8.4 GHz data in the \textit{uv} plane with an elliptical Gaussian with a major axis of $0.25^{\prime\prime} \pm 0.04^{\prime\prime}$, unresolved minor axis ($< 0.1^{\prime\prime}$), and position angle of $92^{\circ} \pm 14^{\circ}$ (degrees east from north). Subtracting this Gaussian in the \textit{uv} plane and imaging the residuals, there is some structure remaining which implies that the source is not fully described as a single Gaussian. The central position of the 2003 emission is near the 2002 position, although not consistent within the formal errors (Table \ref{tab:aconf}). Inspection of the images implies that they likely share a common position, but the more complex morphology of the 2003 image leads to a slight apparent offset.

During the first months of the 2004 A configuration (1408--1488 days after eruption), V445~Pup is
resolved into two distinct regions of emission that are roughly aligned with the major axis of the 2003 elliptical Gaussian fit. 
In the 8.4 GHz observations, the structure of V445~Pup can be fit with two circular Gaussian components separated 
by $0.22^{\prime\prime} \pm 0.02^{\prime\prime}$. No significant motion of the components could be seen over $\sim$4 months of observations.
The north-eastern (NE) component's ICRS 
coordinates are RA = 7h37m56.890s, Dec = $-25^{\circ}56^{\prime}58.80^{\prime\prime}$, and the coordinates of the south-western (SW) component are RA = 7h37m56.875s, Dec = $-25^{\circ}56^{\prime}58.88^{\prime\prime}$.  
The components themselves are marginally resolved, with the NE component's FWHM $= 0.16^{\prime\prime} \pm 0.05^{\prime\prime}$ 
and the SW component measured to have a FWHM $= 0.12^{\prime\prime} \pm 0.03^{\prime\prime}$. The 2004 components are located on either side of the radio emission imaged in 2002/2003, consistent with them moving away from this origin position. The two 2004 components are similar in brightness, with flux densities of $0.7 \pm 0.1$ mJy (NE) and $0.5 \pm 0.1$ mJy (SW).
The NE component gradually fades while the SW component gradually brightens
and remains the only visible component by the end of 2004.

Figure \ref{fig:hires} compares a VLA image produced by stacking the 8.4~GHz observations obtained during the 2004 A configuration
with the near-IR high-resolution image, obtained by \citet{woudt2009} on 2005 Mar 26 using the NAOS/CONICA adaptive optics system of the VLT. To make the near-IR K-band image,
we downloaded several K-band exposures from the European Southern Observatory (ESO) Science Archive Facility (request \#543179),
stacked them to remove cosmic rays and artifacts, and applied a world coordinate system (WCS) using six stars in the image with \emph{Gaia} Data Release 2 (DR2) positions (estimated error on our WCS is $0.02^{\prime\prime}$).  The near-IR K-band emission is a composite of emission lines from warm gas (e.g., \ion{He}{i}) and warm dust continuum emission  \citep{woudt2009}.

From Figure \ref{fig:hires}, it is clear that the radio  emission is emanating from the same bipolar regions of the V445~Pup ejecta that are dominating the near-IR emission. The two images were obtained around the same time (day $\sim$1445 at 8.4 GHz, and day 1605 at near-IR K-band), and to first order, the two radio components are aligned with the peaks of the near-IR emission. \citet{woudt2009} show that the thermal ejecta expand with a range of velocities in the polar (NE--SW) direction, while expansion in the equatorial plane is confined by a dense dust disc. It is the ejecta nearest to the dust disc that is densest and brightest in the near-IR---and also in radio emission. 

We estimate the velocity at which the radio components are traveling apart from one another, assuming they were expelled on $t_0 = 51850$ MJD and V445~Pup is at a distance of 8.2 kpc. In convenient units, where $V$ is the expansion velocity in km~s$^{-1}$ (assumed to be symmetric from a central location), $d$ is in kpc, $t$ is in units of 100 days, and $\theta$ is the angular separation in arcsec:
 \begin{equation}
   V = 8,660\ {\rm km\ s^{-1}}\ \left(\frac{d}{\rm kpc}\right)\ \left(\frac{\theta}{\rm arcsec}\right)\ \left(\frac{\rm 100\ days}{t}\right) 
	\label{eq:expansion_velocity}
\end{equation}
The component separation of $0.22^{\prime\prime}$ on day 1445 implies that the blobs were expanding at 1080 km~s$^{-1}$. If we estimate that both components have FWHM of $0.12^{\prime\prime}$, then the outer portions of the blobs would be separated by $0.34^{\prime\prime}$, implying expansion velocities of 1670 km~s$^{-1}$. 
Given these expansion velocities, it is not surprising that we do not observe significant motion during the 2004 A configuration; between 1408 and 1526 days after eruption, we would only expect the two components to move  by $0.02-0.03^{\prime\prime}$---substantially less than our observational errors (Table \ref{tab:aconf}).
The velocities of 1000--2000 km~s$^{-1}$ implied by the radio imaging are faster than the velocities of the P~Cygni absorption troughs observed in optical spectroscopy a few months after eruption \citep{iijima2008}, but are slower than the fastest expanding ejecta imaged in the near-IR by \cite{woudt2009}.

We note that the high-velocity infrared ``knots" pointed out at the extremities of the bipolar V445~Pup nebula by \citet{woudt2009}\footnote{They are not visible in the K-band image presented in Figure \ref{fig:hires}; this may be because we did not stack all the exposures from 2005 Mar 26, and so our image is shallower than the one published by \citep{woudt2009}. We also did not deconvolve the K-band image as done by Woudt et al.}, seen to expand at a remarkable 8,450 km~s$^{-1}$, would have been easily resolved in our radio images if they were radio bright. The high-velocity knots are not detected in our 2004 A configuration radio image---a clear demonstration that they are not the source of the bulk of the radio emission $\sim$1500 days after eruption. \citet{woudt2009} estimate that the high-velocity knots were ejected 345 days after $t_0$ (a few months before the 2002 A configuration imaging campaign). If the high-velocity knots had been radio-emitting during our 2003 A configuration campaign, they would have been separated by $\sim0.7-0.8^{\prime\prime}$, easily resolvable with our high-resolution radio images. We therefore conclude that it is not the fastest-expanding material in V445~Pup that is dominating the radio emission at any time. Instead, our radio images imply that it is the densest material that is the site of the radio synchrotron emission.

In the 2006 A configuration (day 1925--2015), V445~Pup has substantially faded, and its morphology has reverted to a similar structure as in 2003 (as best we can tell, given the relatively low S/N). We averaged in the \textit{uv}-plane all 8.4 GHz data obtained in this A configuration, and found the data can be well-described
by a single elliptical Gaussian with a major axis of $0.32^{\prime\prime} \pm 0.08^{\prime\prime}$, unresolved minor axis ($\lesssim 0.2^{\prime\prime}$), and position angle of $68^{\circ} \pm 42^{\circ}$ (degrees east from north). The center of this emission region is consistent with the position of the 2003 emission; it is located between the 2004 components.

By the time of the 2007 A configuration, V445~Pup had faded too much for its morphology to be constrained by our VLA images. 

\section{Discussion}

\subsection{Radio Emission from V445~Pup is Synchrotron-Dominated} 
\label{sec:synch}

Historically, radio emission from hydrogen-rich classical novae was thought to be dominated by thermal free-free radiation from expanding ionized ejecta \citep{Seaquist1977,Hjellming1979, seaquist2008}. At early times, while the radio luminosity is increasing, the radio-emitting region is optically thick. As the ejecta expand and drop in density, the radio light curve peaks and turns over, as the radio photosphere recedes through the ejecta and the radio emission transitions to an optically thin state. During the optically thick phase of a thermal-dominated radio light curve, the spectral index $\alpha$ is expected to be equal to 2, as for blackbody emission in the Rayleigh-Jeans long-wavelength limit. When the emitting region is optically thin, the spectral index is flat ($\alpha = -0.1$), as expected for bremsstrahlung emission
\citep[e.g. \S~6.2 in][]{1970ranp.book.....P}. The radio light curve rise, peak, and decay occur on timescales of months to years---much slower than the evolution of optical light curves of novae. 

The other possible source of nova radio emission is synchrotron radiation. The spectral index for optically thin synchrotron emission is set by the energy spectrum of relativistic electrons; if the number of relativistic electrons per unit energy is $N(E) \propto E^{-p}$, then the spectral index is $\alpha = (1-p)/2$. For classic diffusive shock acceleration of relativistic particles, $p = 2-2.5$, and so $\alpha = -0.5$ to $-0.75$ \citep{Bell1978, Blandford1978}. Indeed, $\alpha \approx-0.7$ is commonly observed in synchrotron-emitting SNe and SN remnants. \citep{Chevalier1982, weiler2002}. However, several novae that are strong candidates for synchrotron emitters have unusually shallow spectral indices of $\alpha \approx -0.1$ to $-0.5$ \citep{Taylor1987, Eyres2009, Weston2016, Finzell2018}. This could imply shallower energy spectra for relativistic electrons in novae (compared to e.g., SNe), or that the synchrotron emitting region is inhomogeneous in terms of particle density and magnetic field strength \citep{Vlasov2016}.
Optical depth effects could also flatten the spectrum; in the jets of active galactic nuclei, partially self-absorbed synchrotron emission routinely produces flat or inverted spectra of the ``radio core'', while the more extended transparent regions of the jet display the usual steep spectrum with $\alpha = -0.7$ \citep{1986A&A...168...17E,2019ARA&A..57..467B}.

The radio spectra of V445~Pup from $\sim$1--4.5 years after optical discovery generally show higher flux densities at lower frequencies (Figure~\ref{fig:spectra}; with the exception of a few epochs which appear optically thick because of their inverted spectra). Spectral indices of V445~Pup are in the range $\alpha \approx 0$ to $-1$ (Figure~\ref{fig:full_lc}).
In the first three years of V445~Pup's evolution, its radio spectral index hovers around $\alpha = -0.5$: slightly shallower than that observed for SNe, but well within expectations of synchrotron emission (especially when taking into account the relatively shallow spectral indices observed for other synchrotron-emitting novae). After day $\sim$1000, the spectral index flattens to $\alpha \approx -0.2$.

To further constrain the radio emission mechanism, we determine the brightness temperature ($T_b$; a parameterization of surface brightness). Nova ejecta typically have temperatures $\sim 10^4$ K, as a direct result of the photoionization of the ejecta by the central hot white dwarf \citep{cunningham2015}. We therefore expect the brightness temperature to be $\sim10^4$ K for a nova in the optically-thick thermal phase of its radio evolution. As thermal radio emission transitions to optically thin, the brightness temperature is expected to drop well below $10^4$ K \citep{Weston2016, Finzell2018}. Synchrotron emission, on the other hand, can reach much higher brightness temperatures, up to $\sim 10^{11}$ K \citep{Readhea94}. Therefore, if the nova's radio brightness temperature is substantially in excess of $10^4$ K, this is a promising indication that it is emitting non-thermal radiation.
 
To estimate the brightness temperature, we must constrain the angular diameter of the object ($\theta$) and measure its flux density ($S_\nu$). We can then calculate the brightness temperature using the formula:
\begin{equation}
    \frac{T_b}{\rm K}= 1.36 \times \bigg(\frac{\lambda}{\rm cm}\bigg)^2 \times \bigg(\frac{S_\nu}{\rm mJy}\bigg) \times \bigg(\frac{\theta_a \theta_b}{\rm arcsec^2}\bigg)^{-1}
	\label{eq:brightness_temperature}
\end{equation}
where $\lambda$ is the observing wavelength, $\theta_a$ is the major axis diameter, and $\theta_b$ is the minor axis diameter. We can estimate $\theta$ if we know the speed at which the nova ejecta are expanding ($V$; 
equation~\ref{eq:expansion_velocity}) 
as a function of time 
($t$; \citealt{seaquist2008}). We take $t$ 
as the time since $t_0$ and estimate 
the expansion velocity 
$V = 1600$ km~s$^{-1}$ along the major axis and $V < 800$ km~s$^{-1}$ along the minor axis (inferred from radio imaging; \S \ref{sec:image}).
We use the observed flux densities at 1.43 GHz (21 cm), which place the strongest constraints on the brightness temperature (compared to higher frequencies).

\begin{figure}
	\includegraphics[width=0.5\textwidth]{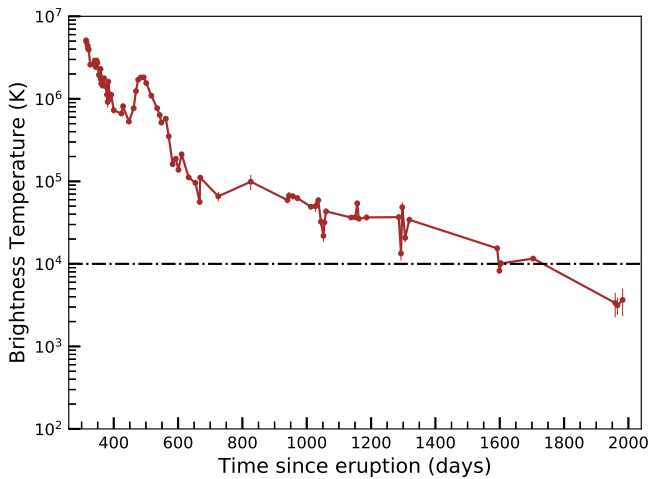}
    \caption{Surface brightness temperature for V445 Pup between 2001 September and 2006 April.}
    \label{fig:effective_temperature}
\end{figure}

As seen in Figure~\ref{fig:effective_temperature}, the brightness temperature of V445~Pup starts high, $\sim 10^7$ K on day $\sim300$, implying that it is undeniably synchrotron emission. The brightness temperature declines with time, to $\sim 10^5$ K by day $\sim600$, and finally to $\sim 10^4$ K by day $\sim2,000$. Despite a relatively low (for synchrotron emission) brightness temperature of few $\times 10^4$ K after $\sim 1000$ days, we conclude that this emission is non-thermal, since the spectral index implies that the emission is optically-thin. If the emission were optically-thin thermal, its brightness temperature should be substantially lower than the $\sim 10^4$ K electron temperature. Therefore, we conclude that the radio light curve of V445~Pup is dominated by synchrotron emission over $\sim$7 years of its evolution.

\subsection{What powers the synchrotron-emitting shocks in V445~Pup?}
\label{sec:origin}

The general consensus is that the source of synchrotron emission in novae is due to acceleration of particles to relativistic speeds in shocks. Depending on the nature of the companion star, these shocks can be either external with pre-existing circumbinary material or internal within the ejecta. Which is it for V445~Pup?

The clearest examples of synchrotron emission in hydrogen-rich classical novae come from embedded novae with red giant companions, where external shocks are produced in the interaction of the nova ejecta and the red giant wind. Examples of this type of synchrotron-emitting shocks are RS Oph \citep{Obrien2006, Rupen2008, Sokoloski2008, Eyres2009}, V745~Sco \citep{Kantharia2016}, and V1535~Sco \citep{linford2017}. In these cases, the radio light curve evolves relatively quickly (over $\sim$weeks), and can be used to trace the radial density profile of the circumbinary material. A late-time thermal radio component is not observed in most cases, indicating that the ejecta are low mass (as expected for the relatively high accretion rates driven by red giant companions; e.g., \citealt{Yaron2005}).

Another way to produce synchrotron emission could be internal shocks within the ejecta. These type of shocks likely dominate in a white dwarf system whose companion is a main sequence star, and hence the binary system is surrounded by less dense circumbinary material. Such shocks form at the interface of two outflows moving at different speeds, 
following a nova eruption. For decades there has been evidence of internal shocks in hydrogen-rich classical novae, based on hard X-ray emission \citep{OBrien1994, Mukai2001, Mukai2008}, and more recently, GeV $\gamma$-ray emission \citep{Ackermann14, Franckowiak2018, Martin2018}. It has also recently been recognized that a significant fraction of these novae show evidence of non-thermal radio emission, manifesting as early bright radio flares. Novae showing evidence of synchrotron emission from internal shocks include QU Vul \citep{Taylor1987}, V959 Mon \citep{chomiuk2014}, V1723 Aql \citep{Weston2016}, V5589 Sgr \citep{weston2016shock}, and V1324 Sco  \citep{Finzell2018}. These flares rise rapidly with time ($\sim$days--weeks), in contrast to expectations for expanding thermal ejecta (rise time $\approx$ months), and have brightness temperatures $\gtrsim 10^5-10^6$ K. In most cases, a second radio maximum becomes visible at late times which is well-described as expanding thermally emitting ejecta.

Is V445~Pup's radio synchrotron emission driven by external or internal shocks? We know little about the companion star due to dust obscuration \citep{ashok2003, woudt2009}, so it is difficult to predict the properties of the circumbinary material surrounding V445~Pup. We can rule out external shocks with relatively spherically-distributed circumbinary material, based on our radio imaging (\S \ref{sec:image}). If the pre-existing material was isotropically distributed, we would expect the synchrotron emission to be brightest at the fastest shocks (see Equation \ref{eq:luminosity}), and so to be easily resolvable (as, e.g., the near-IR high-velocity knots; \citealt{woudt2009}). Instead, the radio emission is confined to much nearer the binary. 

Based on radio imaging of V959~Mon, \citet{chomiuk2014} hypothesize that at the beginning of a nova eruption, a slow outflow concentrated in the equatorial plane of the binary system is generated. A more isotropic fast wind then follows, primarily escaping in the polar directions. The collision between the two distinct flows results in shocks and particle acceleration. Based on near-infrared and radio imaging in the years following the 2000 eruption, V445~Pup exhibits a similar bipolar outflow and an equatorial disc \citep{woudt2009}. Our radio imaging reveals that the synchrotron emission is concentrated near the equatorial plane, even 3--5 years after explosion (\S \ref{sec:image}, Figure \ref{fig:hires}). We also know that the equatorial disc maintains its structure for years following the eruption \citep{woudt2009}. Therefore, interactions between a polar flow and equatorial disc, as proposed for V959~Mon, are the most likely source of the synchrotron emission.

The origin of the equatorial disc, however, is unclear. It may have pre-dated the nova eruption, as mass lost in a previous nova eruption, or from the binary during quiescence (perhaps from the outer Lagrange points during mass transfer; e.g., \citealt{Pejcha2016a}). Indeed, from pre-eruption photometry, \citet{woudt2009} deduced substantial circumbinary dust around V445~Pup in quiescence. The very strong IR signatures of dust early in the eruption led \citet{lynch2001} to surmise that the dust may have pre-dated the eruption, a conclusion also reached by \citet{shimamoto2017infrared} when they measured very large dust masses ($\sim 5 \times 10^{-4}$ M$_{\odot}$) around V445~Pup in 2006.
On the other hand, in at least the case of V959~Mon, an equatorial disc-like structure was ejected during the nova eruption itself \citep{chomiuk2014, linford2015}. It is hypothesized that the equatorial structure was produced by the puffed-up nova envelope, marginally bound to the binary shortly after thermonuclear runaway \citep{chomiuk2014}. As the binary orbited inside the envelope, it transferred energy to help the envelope expand, preferentially in the equatorial direction (i.e., \citealt{Pejcha2016b}). It is worth noting that, unlike in V959~Mon, the equatorial disc is not observed to expand or diffuse during the eruption of V445~Pup; it remains present in imaging until at least 2015, and significant dust obscuration persists around the binary up until the present day (Woudt et al.\ 2021, in prep). This relative constancy of the equatorial disc over $\sim$ two decades may suggest a pre-eruption origin.

\subsection{Understanding the synchrotron-dominated light curve of V445~Pup}
\label{sec:lc_interp}

The radio synchrotron emission from V445~Pup persists for an unprecedentedly long time ($\sim$7 years; Figure \ref{fig:full_lc}). Synchrotron emission in other novae lasts for just a few weeks--months \citep{Eyres2009, Weston2016, weston2016shock, Kantharia2016, linford2017}. One relatively simple way to explain the long synchrotron duration of V445~Pup would be if a wind was launched from the binary and continued blowing for years after eruption, while the equatorial disc retained its structure over this time. Then ongoing interactions between the wind and the disc would continue to power synchrotron luminosity. The standard mechanism for driving prolonged winds in novae is radiation pressure from an $\sim$Eddington luminosity white dwarf, powered by nuclear burning on the white dwarf's surface \citep{Kato1994}. While $\sim$7 years is relatively long for this sustained burning phase, it is not unprecedented amongst hydrogen-rich classical novae \citep{Henze2014}. The larger envelope masses in helium novae lead to expectations of longer durations for the sustained burning phase in systems like V445~Pup ($\sim$few yr -- 10,000 yr; \citealt{Kato2004}). Therefore, a wind prolonged over $\sim$7 years seems reasonable. Unfortunately, we are not able to directly test if a wind in V445~Pup was powered by the nuclear-burning white dwarf, as no X-ray observations of the eruption are available, and the dusty equatorial disc would likely have absorbed the white dwarf supersoft X-ray emission.

To explain the synchrotron emission from V445~Pup, we consider a cartoon model where the white dwarf wind impacts upon the equatorial disc (we assume the disc is not expanding, as it has appeared largely unchanged in imaging over $\sim$15 years post eruption; \citealt{woudt2009}, Woudt et al.\ 2021, in prep.). We approximate that the wind expands at 1600 $\rm km~s^{-1}$, based on our radio imaging (\S \ref{sec:image}), noting that this is the measured velocity in the polar (NE-SW) direction. We take a scenario where a wind of this velocity emanates from the white dwarf isotropically, but is decelerated in the equatorial direction by the disc, creating shocks, accelerating particles, and producing the observed synchrotron emission. We use a simple prescription for synchrotron luminosity, as described in Appendix A and largely taken from \citet{Chevalier1982, Chevalier1998}. We assume a fraction of the post-shock energy density is transferred to energy density of the amplified magnetic field ($U_B = \epsilon_B \rho_{\rm disc} v_{\rm w}^2$) and of relativistic electrons ($U_e = \epsilon_e \rho_{\rm disc} v_{\rm w}^2$). Here $\rho_{\rm disc}$ is the density of the pre-shock material and $v_{\rm w}$ is the velocity of the wind, $\epsilon_B$ and $\epsilon_e$ represent the fraction of the post-shock energy due to amplified magnetic fields and the relativistic electrons, respectively. We note that this is likely an oversimplification if the wind significantly decelerates in the equatorial direction, implying that a reverse shock contributes to 
the shock luminosity \citep{Metzger2014}. A thorough treatment of V445~Pup's shocks and resultant synchrotron luminosity requires a multi-dimensional hydrodynamics simulation, so our goal here is only to draw a simple cartoon.  When the synchrotron emission is optically thin, its luminosity ($L_{\nu}$) is proportional to the synchrotron-emitting volume $V_{\rm synch} \times U_e \times U_B^{(p+1)/4}$. For V445~Pup, the index of the relativistic electron energy spectrum is measured to be $p \approx 2$ (\S \ref{sec:synch}); therefore
\begin{equation}
L_{\nu} \propto V_{\rm synch}\, \epsilon_e\, \epsilon_B^{3/4}\, \rho_{\rm disc}^{7/4}\, v_{\rm w}^{7/2}
\label{eq:luminosity}
\end{equation}
As described in Appendix A, the synchrotron luminosity of V445~Pup can be explained if $\rho_{\rm disc} \approx 10^4$ cm$^{-3}$. 

Higher synchrotron luminosities imply faster shocks and/or higher densities for interaction. Therefore, the variations in synchrotron luminosity observed in V445~Pup's light curve imply variations in the wind velocity or density of the equatorial disc---or in the optical depth. Synchrotron-dominated light curves of radio transients often show an optically-thick rise, where the flux increases as $\tau$ drops, and an optically-thin decline, usually interpreted as a decline in the density of material being shocked and/or a decline in the shock velocity \citep[e.g.,][]{weiler2002}.
V445~Pup is unusual in showing multiple peaks in its light curve (although see the embedded nova V1535 Sco for a similar albeit less dramatic case; \citealt{linford2017}). 

These flares are accompanied by changes in the optical depth. While most of the time the synchrotron emission appears optically thin (flux density monotonically rising toward lower frequency or $\alpha < 0$), on the rise to radio peaks, the radio spectrum flips to be brighter at higher frequencies (Figures \ref{fig:full_lc} and \ref{fig:spectra}). Between days 429 and 447 (on the rise to the brightest flare observed, peaking on day $\sim$460), the flux density increases with frequency all the way up to 22.5 GHz, implying  $\alpha = 0.2$ to  0.4. On days 337--341 (rising to a slightly fainter peak on day $\sim$340), the 1.4 GHz point falls substantially below the power law fit to higher-frequencies, implying that absorption is present but milder than on day 429.  
In both cases, the radio spectrum returns to optically thin at the peak of the radio flare, and thereafter. The observed cyclic change in the radio spectrum of V445~Pup---starting with optically thin emission, then becoming optically-thick, then thin, and then thick and thin again---is unusual and challenging to explain (although again, similar sudden, cyclic changes of $\alpha$ are seen in V1535 Sco; \citealt{linford2017}). With a few rough calculations, we can exclude the possibility that the optical depth is due to synchrotron self-absorption (Appendix B). The optical depth is therefore due to free-free absorption, 
and changes in opacity mean that the ionization state of the absorbing material must be changing. The spectrum never reaches the canonical $\alpha = 2$ for free-free absorption, implying that the emission is only partially optically thick---perhaps because the $\tau \approx 1$ photospheres shrink with frequency as expected in a stellar wind \citep{Panagia1975, Wright1975} or because the covering factor of the absorbing screen is substantially smaller than unity \citep[e.g.,][]{Diaz2018}.

So, what might explain the observed radio flares, and attending changes in free-free opacity? One possibility is that the white dwarf wind is sweeping through the inner parts of the equatorial disc, and encountering density enhancements within it. This would imply that there are at least four regions of enhanced density in the disc, to explain the flares on day $<$300 (we only captured its decline), day 340, day 460, and day $\sim$1300 (Figure \ref{fig:full_lc}). Simulations of equatorial discs produced by mass loss from the outer Lagrangian points of a mass-transferring binary often find structured discs with spiral-like density enhancements \citep{Pejcha2016a, Pejcha2016b}. The flux of V445~Pup rose by a factor of $\sim$10 (from 2 mJy to 22 mJy at 4.9 GHz) between day 430 and 460. According to Equation \ref{eq:luminosity}, this would imply an increase in density of a factor of $\sim$4 to explain this flare.

This interpretation is supported by the rapid declines of the radio flares, which we parameterize as  $S_{\nu}~\propto~t^{\beta}$; the value of $\beta$ can give hints on the source of the synchrotron emission. \citet{Harris2016} performed hydrodynamic simulations of a shock interacting with ``shells" of enhanced density. During impact with a shell, the energy densities in relativistic electrons and magnetic fields increase dramatically, and therefore the radio light curve will peak. However, as the shock propagates to the outer side of the shell, there is no more material to sweep up, particle acceleration ceases, and the energy densities in relativistic electrons and magnetic fields rapidly drop as expected for adiabatic expansion.
 At this point the radio light curves decline rapidly as $L_{\nu}~\propto~t^{-11.5~ \rm to~ -9}$ based on their simulations and analytical relations.
Considering the declines from the radio flares on days 337 and 462 (see Figure~\ref{fig:zoom_lc}), $\beta$ is in the range $-10\lessapprox \beta \lessapprox -9$, consistent with the shell-interaction simulations. 

The other possibility is that the velocity of the wind is variable with time, and synchrotron flares are produced when it is faster. To produce a factor of ten increase in synchrotron luminosity requires a factor of $\sim$2 increase in the wind velocity, according to Equation \ref{eq:luminosity}. Other nova eruptions have been observed to host multiple outflows with a range of velocities, seen in optical spectra as new absorption features and broadening emission lines \citep[e.g.,][]{jack2017study,aydi2019flaring}. \citet{aydi2020direct} observed rapid changes in $\gamma$-ray flux and shock luminosity over the first $\sim$40 days of V906~Car's 2018 nova eruption, and attributed this to changes in outflow velocity with support from optical spectroscopy. However, the flares observed in V906~Car, and indeed most hydrogen-rich novae, occur over substantially shorter periods than the $\sim$few years covered by our V445~Pup radio light curve. Perhaps helium nova eruptions evolve more slowly due to their likely larger envelope masses.
However, this is difficult to test, as there was essentially no spectroscopic monitoring of V445~Pup during the time window 1--3 years after explosion, due to very heavy dust obscuration \citep{ashok2003, lynch2004, iijima2008}, and so it is difficult to constrain if or how the  wind velocity changed during this time. 

In either case---whether it is variations in density or velocity that produce the radio flares---the likely cause of the changes in optical depth is changes in shock luminosity. If much of this shock luminosity is radiated in the ultraviolet band, it can ionize portions of the equatorial disc ahead of it. The higher the shock luminosity, the more ionized material there will be, and the stronger the free-free absorption. 
The free-free absorption later dissipates, either because this material recombines or because it is swept up by the shock. We prefer the latter explanation, because if the disc material has a density  $\sim10^4$ cm$^{-3}$ (Appendix A), the recombination time should be of order a century \citep{ferland2003}: much longer than the $\sim$200 day duration of the radio flare. 




\subsection{Is V445~Pup a SN Ia progenitor system?}

As the only helium nova known, V445~Pup is often singled out as an intriguing candidate for a progenitor system to SNe Ia (e.g., \citealt{Li11, McCully14, Kelly14}). The progenitors of SNe Ia have been constrained with radio continuum observations, where the synchrotron-emitting blast wave formalism of Chevalier and collaborators (i.e., Appendix \ref{sec:synch_model}) is used to place limits on the density and distribution of circumstellar material near the SN site \citep[e.g.,][]{Panagia06, Chomiuk16, Lundqvist20}. 
Tested configurations of circumstellar material are usually over-simplistic, with smooth density profiles and spherical symmetry. Therefore interpretations of radio limits for SNe Ia have substantial associated uncertainty.

The fact that V445~Pup's radio light curve is synchrotron dominated and powered by interaction with an equatorial disc that likely pre-dated the nova eruption  presents a unique opportunity to predict what a radio light curve of a realistic SN Ia might look like. We had originally hoped that we could  simply ``scale up" from nova ejecta energetics to SN ejecta energetics to predict how a SN exploding in a V445~Pup-like progenitor system would appear at radio wavelengths. However, there are two problems with this strategy. First, the synchrotron emission in V445~Pup was likely produced by a prolonged wind that continued to blow and interact with the disc for years---very different from the impulsive yet homologous explosions of SNe. Second, while the equatorial disc apparently withstood the nova eruption of V445~Pup, it is not clear if it would be destroyed by SN Ia ejecta---and on what timescale. 

In the future, hydrodynamic simulations should model the V445~Pup wind/disc system and work to constrain the mass loss rate of the white dwarf wind and the density profile of the equatorial disc, using our radio light curve as a critical constraint on the dynamics. This would inform the ejecta mass and energetics of the only helium nova known, and also recreate the circumstellar environment of one of the most promising SN Ia progenitor systems. 
Once the mass, extent, and morphology of the equatorial disc are better constrained, another hydrodynamic simulation could model the interaction of a SN Ia-like explosion with the disc, and make real predictions for observable signatures at radio, optical, and X-ray wavelengths that might indicate a V445~Pup-like progenitor (see \citealt{Booth2016} for a similar strategy implemented on another SN Ia progenitor candidate, RS~Oph).





\section{Conclusions}

The helium nova V445 Pup is observed at radio frequencies for years following its eruption in late 2000 (Figure~\ref{fig:full_lc}). Steep spectral indices and high brightness temperatures imply that the radio light curve is powered by  synchrotron emission through 2008; we see no evidence of thermal emission from an expanding ionized nova remnant, as seen in many other novae at radio wavelengths. The radio light curve is characterised by at least four re-brightening events, which sometimes are subject to free-free absorption on the rise but revert back to optically-thin synchrotron emission as they fade. Spatially-resolved radio images show that the synchrotron emission is more compact than the thermal ejecta and is confined near the equatorial disc imaged at near IR wavelengths (Figure~\ref{fig:hires}).

We hypothesise that a wind from the white dwarf interacts with the equatorial disc, giving rise to shocks, particle acceleration, and synchrotron emission. 
This model is very similar to the scenario proposed to explain radio synchrotron and GeV $\gamma$-ray emission in hydrogen-rich classical nova V959~Mon, also based on radio imaging \citep{chomiuk2014}. However, the data presented here on V445~Pup present the clearest case to date of  synchrotron emission being associated with a dense equatorial disc in a nova. We note that, based on its high synchrotron luminosity and detection of $\gamma$-rays in other novae \citep{Ackermann14, Franckowiak2018}, V445~Pup was likely a source of substantial GeV $\gamma$-rays, but unfortunately its eruption occurred between the \emph{Compton Gamma-Ray Observatory} and \emph{Fermi Gamma-Ray Space Telescope} missions.

The duration of bright synchrotron emission from V445~Pup ($\sim$7 years) is unprecedented amongst novae. In our model, this implies that the equatorial disc must have maintained its structure throughout these years (a hypothesis directly supported by near IR imaging of V445~Pup; \citealt{woudt2009}). In addition, the wind from the white dwarf must have persisted over these years, which is not surprising if the wind is powered by radiation pressure from sustained burning of helium on the white dwarf's surface \citep{Kato2004}. The re-brightening events in the radio light curve could be produced by changes in wind velocity or density enhancements encountered in the disc, as it is swept up by the wind. Similar fluctuations in shock luminosity have been seen in GeV $\gamma$-rays---albeit over shorter timescales---in the 2018 nova V906~Car \citep{aydi2020direct}.

 



\section*{Acknowledgements}

The authors are grateful to Dr.\ Chelsea E.\ Harris and Dr.\ Ken J.\ Shen for useful discussions. They also thank Prof.~ Michael Rupen, 
Dr.~Amy Mioduszewski and Dr.~Vivek Dhawan for their work in obtaining the VLA observations presented here. 
The National Radio Astronomy Observatory is a facility of the National Science Foundation operated under cooperative agreement by Associated Universities, Inc.

This research was supported by the South African Radio Astronomy Observatory, which is a facility of the National Research Foundation (NRF), an agency of the Department of Science and Innovation. MMN and PAW kindly acknowledge financial support from the University of Cape Town and the NRF.
V.A.R.M.R.\ acknowledges financial support from the Funda\c{c}\~{a}o para a Ci\^encia e a Tecnologia (FCT) in the form of an exploratory project of reference IF/00498/2015/CP1302/CT0001, FCT and the Minist\'erio da Ci\^encia, Tecnologia e Ensino Superior (MCTES) through national funds and when applicable co-funded EU funds under the project UIDB/EEA/50008/2020, and supported by Enabling Green E-science for the Square Kilometre Array Research Infrastructure (ENGAGE-SKA), POCI-01-0145-FEDER-022217, and PHOBOS, POCI-01-0145-FEDER-029932, funded by Programa Operacional Competitividade e Internacionaliza\c{c}\~ao (COMPETE 2020) and FCT, Portugal. L.C.\ and K.V.S.\ acknowledge financial support of NSF awards NSF AST-1751874 \& AST-1907790, NASA grants \emph{Fermi}/80NSSC18K1746 \& \emph{NuSTAR}/80NSSC19K0522, and a Cottrell fellowship of the Research Corporation. J.S. acknowledges support from the Packard Foundation.

\textit{Software}: CASA \citep{McMullin2007}, AIPS \citep{Greisen2003}, difmap \citep{Shepherd1997}, SciPy \citep{Oliphant2007}, Numpy \citep{Van2011}, matplotlib \citep{Hunter2007}.
\\

\section*{DATA AVAILABILITY}
The data presented in this work are available through the VLA archive which can be accessed via https://archive.nrao.edu/archive/advquery.jsp.



\bibliographystyle{mnras}
\bibliography{References} 



\appendix

\section{Simple Model for Synchrotron Emission from Novae}
\label{sec:synch_model}

\citet{Chevalier1982} developed a simple formalism for interpreting synchrotron emission from stellar explosions. The model was developed for application to SNe, where the ejecta blast wave crashes into pre-existing circumstellar material. The shock between the ejecta and the circumstellar material accelerates particles to relativistic speeds (through diffusive shock acceleration; e.g., \citealt{Blandford1978, Bell1978}) and amplifies the magnetic field (through the streaming instability; e.g., \citealt{Bell2004}). The relativistic electrons gyrate along the magnetic field lines to give rise to synchrotron emission. Here we give a few more details of how this model might be applied to V445~Pup.
 
The magnetic field strength in the synchrotron-emitting region is simply $B = \sqrt{8 \pi U_B}$. The energy density of relativistic electrons is spread between electrons with a power-law distribution of energies, $dN(E) = N_0 E^{-p}dE$ where $N_0$ is a constant, and we assume a minimum energy of relativistic electrons equivalent to the rest mass energy of the electron \citep{Chevalier1998}. We take $p=2.1$, as measured from the spectrum of V445~Pup (\S \ref{sec:synch}), noting that some of the equations in \citet{Chevalier1998} are not defined for $p=2.0$.

To convert from a magnetic field strength to physical parameters of the system, we assume a fraction of the post-shock energy density ($\rho_{\rm disc} v_{\rm w}^2$) is converted into amplified magnetic fields and relativistic electrons (as described in \S\ref{sec:lc_interp}). In supernova synchrotron models, conversion factors of $\epsilon_e = \epsilon_B = 0.1$ are often assumed \citep[e.g.][]{chevalier2006, Chomiuk12,cendes2020thirty}. However, recent considerations imply that $\epsilon_B$ could be  much lower  ($\lesssim$0.01, \citealt{kundu2017constraining, Lundqvist20}), and $\epsilon_e$ may also be lower at slower shock velocities \citep[e.g.,][]{sarbadhicary2017}. To roughly estimate the density of material surrounding V445~Pup, we take $\epsilon_e = \epsilon_B = 0.01$.

We estimate $V_{\rm synch} \approx 10^{47}$ cm$^{-3}$, as for a shell that has been expanding for 460 days at 1600 km\,s$^{-1}$, with a thickness of 10\%. If we take a relatively modest flux density for V445~Pup around this time of 3~mJy at an intermediate frequency of 8.4 GHz (consistent with measurements both before and after the day 460 radio flare), this implies that the shock is interacting with material of density, $\sim$3000 cm$^{-3}$. Given that this interaction is probably with the equatorial disc, our estimate of a spherical $V_{\rm synch}$ is over-simplistic. The implied density will increase as the volume filling factor of the synchrotron-emitting material decreases.

\section{Synchrotron Self Absorption in V445~Pup?}
\label{sec:SSA_or_ff}
Could synchrotron self-absorption be the cause of the opacity on the rise to the radio peaks? We can use the flux density at radio peak on day $\sim$460, combined with a few other rough estimates of system parameters, to estimate the opacity due to synchrotron self-absorption ($\tau_{\rm SSA}$). If synchrotron self-absorption is the dominant source of opacity, we would expect $\tau_{\rm SSA} \approx 1$ at radio peak.


Using the same assumptions described in Appendix A, but instead taking
a flux of 12 mJy,
we estimate $B = 0.01$ G.
We can then plug these quantities into the equation for synchrotron self-absorption optical depth (Equation 1 of \citealt{Chevalier1998}),
and find that $\tau_{\rm SSA} \approx 10^{-8}$ at the peak of the day 460 radio flare. As $\tau_{\rm SSA}$ is eight orders of magnitude smaller than unity, we conclude synchrotron self-absorption is not the dominant source of opacity. We note that for $\tau_{\rm SSA} \approx 1$, the emitting volume would need to be $V_{\rm synch} \approx 10^{38}$ cm$^{-3}$, a factor of a billion smaller than our estimate in Appendix A---but the emitting region is directly constrained by our radio imaging (\S \ref{sec:image}), and exclude this possibility.

\section{VLA observations of V445 Pup, measured flux densities and spectral indices}
\label{sec:observations_and_fluxes}
\begin{table*}
 \centering
   \centering
    \caption{Log of VLA observations of V445~Pup.}
 \label{tab:continued_VLA_obs}
    \begin{threeparttable}
\begin{tabular}{lccccccccr}
\hline
Observation & $t$ & $t-t_0$ & Configuration & \multicolumn{6}{c}{Observation time on target (min)}\\
Date & (MJD) & (Days) & &1.43 GHz & 4.86 GHz & 8.46 GHz& 14.94 GHz & 22.46 GHz & 43.34 GHz \\ 
\hline 
2001 Oct 04 & 52187 &337 & C\&D & 8.9 & 6.7 & 6.7&6.7 &6.7 & 6.2 \\
2001 Oct 08 & 52191 &341 & C\&D& 3.1& 4.7 &4.7& 3.7 &5.2 &5.4 \\
2001 Oct 10 & 52193 &343 & C\&D&3.6 & 4.2& 4.2 & 5.1&5.2 &5.6\\
2001 Oct 12 & 52195 &345 & D & 2.6 & 4.1& 4.2& 5.1 &5.2 & 5.6 \\
2001 Oct 14 & 52197 &347 & D &4.4  & 4.2& 4.2& 5.7 &6.9 &6.4 \\
2001 Oct 17 & 52200 &350& D &10.2 &4.2& 4.2&6.2 & 7.6& ...\\
2001 Oct 22 & 52204 &354 & D &2.9 & 2.7 & 2.4& 5.2 &... &4.7 \\
2001 Oct 24 & 52207 &357 & D &4.9 & 2.7 & 2.8 & 6.9 &7.6 &... \\
2001 Oct 26 & 52209 &359 & D &5.4 & 5.7& 7.4& 7.7&7.6 &... \\
2001 Oct 27 & 52210 &360& D &7.3 & 5.6 &7.4&7.7 &7.7 &... \\
2001 Oct 29 & 52212 & 362 & D &4.4 &4.7 & 4.7&6.9 &7.4 & ... \\
2001 Nov 02 & 52216 & 366 & D &7.6 &8.4 & 8.6&7.6 &7.7 &... \\
2001 Nov 04 & 52218 & 368 & D & 7.8&8.4 & 24.6&9.2 & ...& ... \\
2001 Nov 06 & 52220 &370 & D &5.9 &5.3 & 5.2& 6.2& ...& ...\\
2001 Nov 07 & 52221 &371 & D &5.9 &4.7 & 4.2&5.8 &6.6 &...  \\
2001 Nov 11 & 52224 &374 & D & 6.1&... & 4.2&5.3 & 6.2& 5.7\\
2001 Nov 14 & 52227 & 377 & D &... & 3.9 &3.7 &... &... &... \\
2001 Nov 15 & 52229 & 379 & D &4.1 &4.2 &2.2 &5.1 &... &...\\
2001 Nov 16 & 52230 & 380 & D & ... & 5.2 & 4.9& 6.2&6.9 &... \\
2001 Nov 18 & 52231 &381 & D &4.2& 5.1& 3.1 &6.2 &6.9 & ...\\
2001 Nov 20 & 52233 &383& D &4.6 &4.2 &4.2&6.0 & 6.9&6.2 \\
2001 Nov 23 & 52236 &386 & D &... &4.9 &... &... &... &... \\
2001 Nov 24 & 52237 &387& D &... &4.4 & 5.1&5.7 &7.1 & ... \\
2001 Nov 26 & 52240 &390& D &8.1 &4.2 &... &6.4 &... &... \\
2001 Nov 29 & 52242 &392 & D & 5.1&5.1& 5.4&6.1 & 7.1&... \\
2001 Dec 07 & 52250 &400 & D &8.2 &4.2 &4.2&6.2 &... &... \\
2001 Dec 09 & 52252 & 402 & D &... &... &5.2 &6.2 &... &... \\
2001 Dec 31 & 52274 &424& D &5.3 &4.2 &4.1&6.2 &... &... \\
2002 Jan 05 & 52279 & 429& D &7.6 & 4.2 & 4.1&6.1 &6.7 &6.2  \\
2002 Jan 12 & 52286 &436 & D & ...& 4.1 & 4.2&4.7 &... &...  \\
2002 Jan 13 & 52287 &437 & D & ...& 4.1& ... &5.6& ...& ...\\
2002 Jan 23 &52297 & 447 & A & 6.4& 5.9& 7.6 &10.6& ...& ...\\
2002 Feb 07& 52312 &462& A &4.2 &3.2 &9.1 &4.7 & ...&... \\
2002 Feb 14& 52319 & 469& A &1.9 &4.6 & 10.9 &4.7 &... &... \\
2002 Feb 21& 52326 & 476 &A &4.1 &2.3 &6.2 &4.7 & ...&... \\
2002 Mar 01& 52334 & 484 & A &3.9 &3.1 &7.6 &3.7 &... &... \\
2002 Mar 06& 52339 & 489  & A &... &... &4.7 &... &... &... \\
2002 Mar 10& 52343 & 493 & A &4.6 & 4.6 &9.1 &3.7 &4.2 &... \\
2002 Mar 18& 52351 & 501 & A &2.6 &3.2 &4.6 &3.2 & 3.2& ...\\
2002 Apr 03& 52367 & 517  & A &3.2 &5.2 &6.1 &3.6 &3.2 & ...\\
2002 Apr 21& 52385 & 535 & A &3.2 &5.2 &6.1 &... &... &... \\
2002 Apr 29& 52393 & 543 & A &6.6 &5.2 &15.3& 5.9 &... &... \\
2002 May 04& 52398 & 548& A &6.1 &5.2 & ... &6.1 &... &... \\
2002 May 17& 52412 & 562 & A\&B &3.3 &4.2 &... &... &... &...\\
2002 May 26& 52421 & 571& A\&B &3.2 &3.1 &... & ...&... & ...\\
2002 Jun 07& 52433 & 583& A\&B &4.0 & 3.2& ...& ...& ...& ...\\
2002 Jun 17& 52443 & 593 & B &3.6 & 3.2& ...& ... &... &... \\
2002 Jun 25& 52451 & 601& B &3.4 &3.2 &3.1 &.. &... & ...\\
2002 Jul 05& 52461 & 611 & B &3.4 &3.2 &3.1 & ...&... &... \\
2002 Jul 19& 52475 &625 & B &... &... &16.3 &... &... &... \\
2002 Jul 27& 52483 &633 & B &4.1 &3.2 &4.7 &... &... &... \\
2002 Aug 17 & 52504 & 654& B &4.72 &6.2 &10.6 &... &... &... \\
2002 Aug 30 & 52517 & 667 & B &10.6 &14.0 &13.6 &... & ...&... \\
2002 Sep 01 & 52519 & 669 & B &6.6 &8.1 &7.9 &... &... & ...\\
2002 Oct 04 & 52552& 702 & B & ...& 7.1 &7.6 &... & ...&... \\
2002 Oct 28 & 52575& 725 & B &4.9 & 4.1 &4.2 &... &... & ...\\
2002 Oct 31 & 52579 & 729 & B &... &... &4.1 &... & ...& 4.7 \\
2002 Dec 09 & 52617 & 767 & C & ...& 8.2 &8.4 &... & ...& ...\\
2002 Dec 21 &52629  & 779 & C &...&6.2 &6.1 &... &... &... \\
2002 Dec 29 & 52637 & 787& C & ...&6.1 &6.1 & ... & ...&... \\
2003 Jan 03 & 52642 &792& C &... &8.1 &8.1 &... &... &... \\
2003 Jan 06 & 52645 &795 & C &... &8.1 &8.2&... &... & ... \\
2003 Jan 08 & 52647 &797 & C &... &8.1 &8.2 &... &... & ... \\

\hline 
\end{tabular}
\end{threeparttable}
\begin{tablenotes}
\item `...' indicates no observations for this epoch at that frequency 
\end{tablenotes}
\end{table*}

\begin{table*}
    \centering
    \contcaption{Log of VLA observations of V445~Pup.}
\begin{tabular}{lccccccccr}
\hline
Observation & $t$ & $t-t_0$ & Configuration & \multicolumn{6}{c}{Observation time on target (min)}\\
Date & (MJD) & (Days) & &1.43 GHz & 4.86 GHz & 8.46 GHz& 14.94 GHz & 22.46 GHz & 43.34 GHz \\ 
\hline 
2003 Jan 14 & 52653 &803 & C &... & 9.2&9.2 &... &... & ... \\
2003 Jan 20 & 52659 &809 & C &... &9.1 &9.1 &... &... & ... \\
2003 Jan 29 & 52668 &818 & C &... &9.7 &9.0 &... &... & ... \\
2003 Feb 06 &52676 & 826 & D & 8.6 &9.7 &9.1& ...& ...&...\\
2003 Feb 23 & 52693 & 843 & D & ... &9.7 &9.1& ...& ...&... \\
2003 Mar 04 &52702 & 852 & D & ... &10.4 &9.9 & ...& ...&... \\
2003 Mar 11 &52709 & 859 & D & ... &10.3 &12.2  & ...& ...&... \\
2003 Mar 13 &52711 & 861 & D & ... &16.4 &22.1  & ...& ...&... \\
2003 Mar 19 & 52717 & 867 & D & ... &10.4 &9.2  & ...& ...&... \\
2003 Mar 26 & 52724 & 874 & D & ... &10.6 &9.2  & ...& ...&... \\
2003 Apr 18 &52748 & 898  & D & ... &15.9 &12.2  & ...& ...&... \\
2003 May 30 &52790  &940& A &8.8 &6.9 &6.2 &... &... & ... \\
2003 Jun 04 &52795 &945 &A &8.6 &7.0 & ...& ...&... & ...\\ 
2003 Jun 16 &52807 &957& A& 8.8 &7.7 &... &... &... &... \\ 
2003 Jun 30 &52821 &971&A &8.9 &7.7 &... &... &... &... \\ 
2003 Jul 18 & 52839 & 989 &A &... &... &15.2 &... & ...&... \\
2003 Jul 21 &52842 & 992 &A &... &15.2 &... &... &... &... \\ 
2003 Jul 25 &52846 &996 &A & ...& 16.2&... & ...&... & ...\\ 
2003 Jul 31 &52856 & 1002 &A &... &... &9.6 &... &... &... \\
2003 Aug 11 &52863 &1013 & A &8.6 &9.2 &11.1 &... & ...&... \\
2003 Aug 25 & 52877 &1027 & A &6.9 &9.1 &10.7 &...& ...& ...\\
2003 Sep 03 & 52886& 1036 & A &7.3 & 9.1&18.2 &... & ...& ...\\
2003 Sep 11 & 52894 & 1044 & A &10.9 & 10.1&22.4 &... &... &... \\
2003 Sep 19 &52902 & 1052 & A & 10.9& ...& 24.4&... & ...& ...\\
2003 Sep  &52905 & 1055 & A &8.6 & 8.1 & 10.7&... & ...& ...\\
2003 Sep 27 & 52910 & 1060 & A\&B &10.8 & 10.1&... &... & ...&... \\
2003 Sep 29 & 52912 & 1062  & A\&B & ...& 10.6&11.6 &... &... & ...\\
2003 Oct 08 & 52921 & 1071 &A\& B & ...&12.2 &... & ...& ...&...\\
2003 Oct 21 & 52934 & 1084 &B & ...&10.2 &14.2 & ...& ...&...\\
2003 Nov 02 &52946 & 1096 & B &... & 8.2& 8.3&... &\\
003 Nov 09 &52953 &1103 & B & ...&8.7 &18.4 &... &... & ...\\
2003 Nov 18 &52962 &1112  & B &... &... &14.4 &... &... & ...\\
2003 Nov 24 &52967 & 1117 & B &... &8.7 &10.2 &... & ...&... \\
2003 Dec 15 &52988 &1138 & B &5.9 &9.1 &12.2 &... &... &... \\
2003 Dec 28 &53001  &1151 & B & 5.9 & 9.1 & 12.2& ...& ...& ...\\
2004 Jan 03 & 53007 &1157 & B &5.1 & 9.1& 12.2& ...&... &...\\
2004 Jan 08 & 53012 &1162 & B & 5.7& 9.2 &11.7 &... &... &... \\
2004 Jan 14 & 53018 &1168 &B & ...&9.2 &11.5 &... &... & ...\\
2004 Jan 23 & 53027 &1177 & B\&C &... &6.2 &6.4 &... &... &... \\
2004 Feb 01 & 53036 &1186 & B\&C &5.9 & 6.2& 6.2& & & \\
2004 Feb 02 & 53037&1187 & B\&C & ...& ...& 4.1 &10.3 &10.2 &10.7 \\
2004 Feb 12 & 53047 &1197 & B\&C &... & 4.2&4.1& ...&9.0 &10.7 \\
2004 Feb 24 & 53059 & 1209 & B\&C & ...& 4.2 & 4.1& ...&11.9 & ...\\
2004 Mar 02 & 53066 & 1216 & B\&C&... & 4.2 &4.1 &... &12.3 &... \\
2004 Mar 10 & 53074 & 1224 &C & ...& 57.4& ...& ...& ...&... \\
2004 Mar 20 & 53084 & 1234 & C & ... & 4.2 & 8.1& ...&...&...\\
2004 Mar 22 & 53086 & 1236 & C& ...&... &6.2 & ...&... &... \\
2004 Apr 01 & 53096 & 1246 & C&...&... & 8.2& ...&... &... \\
2004 Apr 07 & 53102 & 1252 & C&... &4.2 & 8.1 &... &13.6 &... \\
2004 Apr 15 & 53111 & 1261 & C& ...&4.2 & 4.8 & ...&... &... \\
2004 May 03 & 53129 & 1279 & C&... & 5.2 & 8.1& ...& ...& ...\\
2004 May 10 & 53136 & 1286 & C& 10.6& 10.6& 9.9&... & ...& ...\\
2004 May 17 & 53143 & 1293 & C\& D& 10.6& 8.9& 10.1& ...& ...&... \\
2004 May 22 & 53147 & 1297 & C\& D& 7.4 &... &... &... &... &... \\
2004 May 30 & 53156 & 1306& C\& D & 6.8& 5.2&5.1 &... & ...&...\\
2004 Jun 12 & 53169 & 1319 &C\& D &6.0 & 8.9&10.1 &... &... & ...\\
2004 Jun 25 & 53182 & 1332& D &... &8.9 &10.1 & ...&... &... \\
2004 Jul 01 & 53188 & 1338 & D&... &8.6 & 10.1& ...& ...&... \\
2004 Jul 10 & 53197 & 1347 &D &... &... & 10.1& 20.4&... & ...\\
2004 Jul 17 & 53204 & 1354 &D & ...& 8.9& 10.1&19.2 &... &... \\
2004 Jul 23 & 53210 & 1360 & D&... & 8.6&6.9 &.... & ...& ...\\
2004 Aug 07 & 53225 &1375 &D & ...& 8.9& 10.1&12.7 &... & ...\\
2004 Aug 21 & 53239& 1389 & D&... & 9.2 & 10.2 &10.2 &... &... \\

\hline 
\end{tabular}
\begin{tablenotes}
\item `...' indicates no observations for this epoch at that frequency 
\end{tablenotes}
\end{table*}

\begin{table*}
    \centering
    \contcaption{Log of VLA observations of V445~Pup.}
\begin{tabular}{lccccccccr}
\hline
Observation & $t$ & $t-t_0$ & Configuration & \multicolumn{6}{c}{Observation time on target (min)}\\
Date & (MJD) & (Days) & &1.43 GHz & 4.86 GHz & 8.46 GHz& 14.94 GHz & 22.46 GHz & 43.34 GHz \\ 
\hline
2004 Sep 09 & 53258 & 1408 & A&... &9.2 & 10.2& 10.2&... &... \\
2004 Sep 23 & 53272 & 1422 & A& ...& 8.6& 12.2& ...& ...&... \\
2004 Oct 01 & 53280 & 1430 & A& ...&... & 18.2& ...&... &... \\
2004 Oct 08 & 53287 & 1437 &A & ...& ...& 17.4&... &... &... \\
2004 Oct 16 & 53295 & 1445 &A & ...&... & 15.2& ...&... &... \\
2004 Oct 31 & 53310 & 1460 & A& ...& 9.1& 9.2&... &... & ...\\
2004 Nov 03 & 53312 & 1462 & A& ...& 8.2&18.2 & ...&... &... \\
2004 Nov 07 & 53317 & 1467 &A & ...&... &14.1 &... & ...&... \\
2004 Nov 12 & 53321 & 1471 & A& ...& ...&21.1 & ...& ...&... \\
2004 Nov 20 & 53329 & 1479 &A & ...&... &20.8 &... & ...&... \\
2004 Nov 29 & 53338 & 1488 & A& ...& ...& 21.2& ...& ...&... \\
2004 Dec 22 & 53361 & 1511 & A& ...&... & 14.6&... &... &... \\
2004 Dec 31 & 53370 & 1520 & A& ...& ...&15.2 &... &... &... \\
2005 Jan 06 & 53376 & 1526 & A& ...& ...& 13.6& ...&... &... \\
2005 Jan 16 & 53386 & 1536 & A\& B &... & ...&10.2 &... &... & ...\\
2005 Jan 24 & 53394 & 1544  & A\& B & ...&... &8.6 & ...& ...& ...\\
2005 Jan 29 & 53399 & 1549  & A\& B & ...& 12.1&11.8 & ...& ...&... \\
2005 Feb 03 & 53404 & 1554  &A\& B &... & ...& 7.1& ...& ...& ...\\
2005 Feb 26 & 53427 & 1577  & B& ...& 8.6& 9.2& ...&... &... \\
2005 Mar 13 & 53442 & 1592 &B & 8.9& 8.1& 10.3&... &... &... \\
2005 Mar 20 & 53449 & 1599 & B&8.8 &8.1 & 10.4& ...&... &... \\
2005 Mar 25 & 53454 & 1604 &B &8.9 & 8.1& 10.2& ...& ...&... \\
2005 Apr 04 & 53464 & 1614 & B& ...&37.7 & ...&... &... &... \\
2005 Apr 14 & 53475 & 1625 & B& ...& 6.6& 8.2&... & ...& ...\\
2005 Apr 26 & 53486 & 1636 & B& ...& 6.4& 9.1& ...&... &... \\
2005 May 06 & 53497 & 1647 &B &... &5.2 &9.0 & ...&... &... \\
2005 May 14 & 53504 & 1654 &B & ...& 5.1& 10.2& ...&... & ...\\
2005 May 29 & 53520 & 1670 & B& ...&22.4&... & ...&... &... \\
2005 Jun 03 & 53525 & 1675 &B & ...& 19.9&... &... &... & ...\\
2005 Jun 12 & 53534 & 1684 & B\&C & ...& ...&10.7 &... &... & ...\\
2005 Jun 18 & 53540 & 1690 & B\&C&... &... &15.4 & ...& ...&... \\
2005 Jun 23 & 53545 & 1695 & B\&C & ...& 13.1 & 12.2  &... &... & ...\\
2005 Jul 02 & 53554 & 1704 & B\&C& 11.4 & 13.1 & 12.2&...&...&... \\
2005 Jul 09 & 53561 & 1711 & C& ...&... &4.2 &... &... &... \\
2005 Jul 14 & 53566 & 1716 & C&... & 10.9&... &... &... &... \\
2005 Jul 23 & 53575 & 1725  & C& ...&10.1 &14.2 &... &... &... \\
2005 Aug 02 & 53585 & 1735  & C& ...&... &17.2 &... &... &... \\
2005 Aug 06 & 53589 & 1739  & C&... &... &6.2 &... &... &... \\
2005 Aug 27 & 53610 & 1760  & C& ...&20.2 &12.2 & ...& ...&... \\
2005 Sep 05 & 53619 & 1769  & C& ...&20.4 &16.7 &... &... &... \\
2005 Sep 25 & 53639 & 1789  & C& ...& 22.2&... &... &... &... \\
2005 Sep 27 & 53641 &1791 &C &... &12.2 & 18.2& ...&... &... \\
2005 Oct 01 & 53645 & 1795 &C & ...& 20.4&16.7 &... &... &... \\
2005 Oct 05 & 53649 & 1799 & C\&D &... &... &14.4 & ...&... &... \\
2005 Oct 16 & 53660 & 1810 & C\&D & ...&... & 10.4&... &18.7 &... \\
2005 Oct 23 & 53666 & 1816 &C\&D &... & 61.6 &14.2 &... &... &... \\
2005 Oct 31 & 53675 & 1825 & C\&D&... &... &18.4 &... &... &... \\
2005 Nov 08 & 53683 & 1833 &D &... & 9.1 & 9.2& ...&... & ...\\
2005 Nov 15 & 53690 & 1840 & D&... & ...& 9.2&... &... &... \\
2005 Nov 21 & 53696 & 1846 & D&... &... & 21.2&... &... & ...\\
2005 Nov 29 & 53703 & 1853 &D & ...& ...& 17.4& ...&... &... \\
2005 Dec 01 & 53705 & 1855 & D&... & 13.2& 15.6& ...&... &... \\
2005 Dec 11 & 53715 & 1865 & D& ...&13.1 &15.6 &... & ...& ...\\
2005 Dec 20 & 53724 & 1874 &D &... &13.1 &15.2 &... &... &... \\
2006 Jan 13 & 53748 & 1898 &D & ...& ...&14.9 & ...&... &... \\
2006 Jan 15 & 53750 & 1900 & D& ...&... &18.2 & ...& ...&... \\
2006 Jan 22 & 53757 & 1907 & D& ...& 9.2&... & ...& ...& ...\\
2006 Feb 09 & 53775 & 1925 & A& ...& ...&16.2& ...& ...& ...\\
2006 Mar 15 & 53809 & 1959 & A& 9.6 & ...& 12.2& ...&... &... \\
2006 Mar 22 & 53816 & 1966 & A& 9.5 & ...& ...& ...&... &... \\
2006 April 07 & 53832 & 1982 & A& 8.9 & ...& ...& ...&... &... \\
2006 May 08 & 53863 & 2013 &A & ...&... & 20.2&... &... &... \\
2006 May 10 & 53865 & 2015 &A &... & ...& 14.2& ...&...&... \\
\hline 
\end{tabular}
\begin{tablenotes}
\item `...' indicates no observations for this epoch at that frequency 
\end{tablenotes}
\end{table*}

\begin{table*}
    \centering
    \contcaption{Log of VLA observations of V445~Pup.}
\begin{tabular}{lccccccccr}
\hline
Observation & $t$ & $t-t_0$ & Configuration & \multicolumn{6}{c}{Observation time on target (min)}\\
Date & (MJD) & (Days) & &1.43 GHz & 4.86 GHz & 8.46 GHz& 14.94 GHz & 22.46 GHz & 43.34 GHz \\ 
\hline

2006 Aug 05 & 53953 & 2103 & B&... &... &24.6 &... &... &... \\
2006 Aug 09 & 53957 & 2107 & B& ...& 24.4&... &... &... &... \\
2006 Aug 31 & 53979 & 2129 & B&... &20.6 &... &... & ...& ...\\
2006 Sep 11 & 53990 & 2140 &B & ...&38.3 &... &... &... & ...\\
2007 Jan 23 & 54123 & 2273 & C\&D &... &... &29.3 &... & ...& ...\\
2007 Sep 30 & 54374& 2524 & A\&B&... &... & 36.0& ...& ...&... \\
2008 Jan 17 & 54482 & 2632 & B& ...& ...&28.0 & ...& ...&... \\
2008 Mar 28 & 54554 & 2704 &C & ...& ...& 10.4& ...& ...&... \\

\hline 
\end{tabular}
\begin{tablenotes}
\item `...' indicates no observations for this epoch at that frequency 
\end{tablenotes}
\end{table*}

\begin{table*}
    \centering
    \caption{Flux densities and spectral indices of V445~Pup}
 \label{tab:continued_VLA_fluxes}
\begin{tabular}{lcccccccr}
\hline
$t$ & $t-t_0$ &\multicolumn{6}{c}{Radio flux densities (mJy)}& $\alpha$\\
 (MJD) & (Days) &1.43 GHz & 4.86 GHz & 8.46 GHz& 14.94 GHz & 22.46 GHz & 43.34 GHz&\\ 
\hline 
52187& 337 & 12.62 $\pm$ 0.67 & 13.85 $\pm$ 0.71& 12.01 $\pm$ 0.62& 8.69 $\pm$ 0.91& 5.98 $\pm$ 0.88 & 5.10 $\pm$ 1.21 & $-0.46~ \pm~ 0.08$ \\
52191 &341 & 14.26 $\pm$ 0.76& 14.01 $\pm$ 0.73& 11.33 $\pm$ 0.60& 7.54 $\pm$ 1.17&8.61 $\pm$ 1.26 &4.30 $\pm$ 1.18& $-0.43~\pm~ 0.07$\\
52193 & 343 & 13.54 $\pm$ 0.80 & 12.48 $\pm$ 0.66 & 9.29 $\pm$ 0.52 & 7.27 $\pm$ 0.88 & 5.13 $\pm$ 1.05& 2.49 $\pm$ 0.78& $-0.58 ~\pm~ 0.06$ \\
52195 & 345 & 12.15 $\pm$ 0.85 & 11.46 $\pm$ 0.63 & 8.67 $\pm$ 0.49&5.83 $\pm$ 0.71&5.66 $\pm$ 0.75 & 7.00 $\pm$ 1.22 & $-0.24 ~\pm~ 0.07$\\
52197 & 347 & 14.75 $\pm$ 1.21& 10.4 $\pm$ 0.57 & 7.64 $\pm$ 0.42& 7.93 $\pm$ 0.87& 7.79 $\pm$ 0.92&7.21 $\pm$ 1.16 & $-0.25 ~\pm~ 0.06$\\
52200 & 350 & 13.96 $\pm$ 1.18 & 8.81 $\pm$ 0.50 & 8.45 $\pm$ 0.46& 8.31 $\pm$ 0.91 &7.36 $\pm$ 1.0 &... &$-0.23 ~\pm~ 0.06$\\
52204 & 354 & 10.30 $\pm$ 1.31& 8.83 $\pm$ 0.79 &7.56 $\pm$ 0.46&5.82 $\pm$ 0.66&... & 3.14 $\pm$ 0.86& $-0.27 ~\pm~ 0.07$\\
52207 & 357 & 10.20 $\pm$ 0.93 &7.37 $\pm$ 0.41&7.30 $\pm$ 0.44 &6.18 $\pm$ 0.68 &4.2 $\pm$ 0.57&... & $-0.24~ \pm~ 0.02$\\
52209 & 359 & 12.5 $\pm$ 1.18 & 8.2 $\pm$ 0.44 & 7.17 $\pm$ 0.37 &5.69 $\pm$ 0.60 &5.64 $\pm$ 0.72 &... & $-0.30 ~\pm~ 0.02$\\
52210 & 360 & 9.42 $\pm$ 0.84 &7.23 $\pm$ 0.40 &6.72 $\pm$ 0.36& 4.52 $\pm$ 0.58 & 3.71 $\pm$ 0.69 & ...&$-0.26 ~\pm~ 0.07$\\
52212 & 362 & 8.42 $\pm$ 0.45 & 7.52 $\pm$ 0.41 & 6.62 $\pm$ 0.38 & 4.30 $\pm$ 0.51 & 3.94 $\pm$ 0.72 &...&$-0.20~\pm~ 0.07$\\
52216 & 366 & 8.14 $\pm$ 0.44 & 6.99 $\pm$ 0.37 &5.29 $\pm$ 0.30 & 3.62 $\pm$ 0.50 & 1.89 $\pm$ 0.41 & ... & $-0.65~\pm~ 0.13$ \\
52218 & 368 & 8.29 $\pm$ 0.42 & 6.33 $\pm$ 0.35 & 4.75 $\pm$ 0.25 & 3.95 $\pm$ 0.52 & ...& ...&  $-0.30~\pm~0.05$\\
52220& 370& 10.26 $\pm$ 0.61 & 5.66 $\pm$ 0.35 & 4.28 $\pm$ 0.26 & 3.16 $\pm$ 0.49& ...& ...& $-0.49~\pm~ 0.01$\\
52221 & 371 & 8.56 $\pm$ 0.45 & 5.52 $\pm$ 0.33& 4.28 $\pm$ 0.26& 3.51 $\pm$ 0.40& 2.53 $\pm$ 0.54 &...& $-0.39~\pm~0.02$\\
52224 & 374 & 8.39 $\pm$ 0.42 & ... & 4.48 $\pm$ 0.28 & 4.57 $\pm$ 0.65 & 2.96 $\pm$ 0.77 & 2.89 $\pm$ 0.477 &$-0.32~\pm~0.03$\\
 52227 & 377 & &...& 4.81 $\pm$ 0.037 &4.29 $\pm$ 0.29 &... & ...& ...\\
52229 & 379 &6.79 $\pm$ 0.93& 4.68 $\pm$ 0.29 &4.71 $\pm$ 0.35 &4.89 $\pm$ 0.83 & ...&...& $-0.15~\pm~ 0.09$\\
52230& 380 &...& 4.5 $\pm$ 0.28 & 5.01 $\pm$ 0.29 & 4.80 $\pm$ 0.60 &  3.72 $\pm$ 0.64 & ...& $-0.22~\pm~0.13$ \\
52231& 381 &5.61 $\pm$ 0.81& 4.81 $\pm$ 0.30 & 5.25 $\pm$ 0.29& 4.38 $\pm$ 0.54 & 3.88 $\pm$ 0.59 & ...& $-0.10~\pm~0.06$\\ 
52233& 383 &9.99 $\pm$ 0.58 & 5.10 $\pm$ 0.33 & 4.66 $\pm$ 0.26 & 3.65 $\pm$ 0.47 & 3.15 $\pm$ 0.48 & 2.14 $\pm$ 0.63& $-0.43~\pm~0.04$\\ 
52236 & 386 & ...& 4.95 $\pm$ 0.41 &  ...& ...& ...& ... & ...\\
52237& 387 & ...&4.49 $\pm$ 0.33 & 4.01 $\pm$ 0.26 &3.15 $\pm$ 0.44 & 2.39 $\pm$ 0.52 & ...&$-0.34~\pm~0.08$ \\ 
52240 & 390 & 6.39 $\pm$ 0.37 & 4.51 $\pm$ 0.32 & ... & 2.35 $\pm$ 0.51 & ...& ... &  $-0.33 ~\pm~ 0.08$\\    
52242 & 392 & 12.38 $\pm$ 1.08 & 4.10 $\pm$ 0.29 & 3.46 $\pm$ 0.20 & 2.30 $\pm$ 0.33 & 2.59 $\pm$ 0.72 & ...& $-0.43 ~\pm~0.03$\\   
52250 & 400 & 4.91 $\pm$ 0.47 & 3.30 $\pm$ 0.30 & 2.49 $\pm$ 0.26 & 1.97 $\pm$ 0.47 & ... & ...&  $-0.38~\pm~0.03$\\ 
52252 & 402 & ...& ... &2.16 $\pm$ 0.22 &2 .30 $\pm$ 0.40 & ... & ...& ...\\   
52274 & 424 & 5.03 $\pm$ 0.27 & 2.02 $\pm$ 0.23 & 1.71 $\pm$ 0.18 & 3.09 $\pm$ 0.71 & ...& ... & $-0.54~\pm~0.16$\\ 
52279 & 429 & 6.31 $\pm$ 0.57 & 2.02 $\pm$ 0.17 & 2.65 $\pm$ 0.19 & 3.12 $\pm$ 0.42 & 3.77 $\pm$ 0.57 & 3.20 $\pm$ 1.15 & 0.36 $\pm$ 0.06\\   
52286 & 436 & ...&3.91 $\pm$ 0.33 & 4.67 $\pm $ 0.33 & 4.36 $\pm$ 0.78 & ... & ... &  0.19 $\pm$ 0.15 \\ 
52287& 437 &...& 3.61 $\pm$ 0.30 &...& 2.80 $\pm$ 0.62 & ...& ... & ...\\
52297& 447 &4.46 $\pm$ 0.29 & 4.93 $\pm$ 0.31 &9.38 $\pm$ 0.52 & 14.01 $\pm$ 1.52 & ...& ... & 0.44 $\pm$ 0.18\\
52312 & 462 & 6.91 $\pm$ 0.36 & 21.92 $\pm$ 1.12 & 17.68 $\pm$ 0.89 & 11.39 $\pm$ 1.30 & ...& ...& $-0.51~\pm~0.13$\\   
52319 & 469 & 11.49 $\pm$ 0.63 & 15.53 $\pm$ 0.83 & 13.02 $\pm$ 0.67 & 4.35 $\pm $ 0.95 & ...& ...& $-0.54~\pm~0.44$ \\
52326 & 476 & 16.27 $\pm$ 0.83 & 14.09 $\pm$ 0.74 & 10.38 $\pm$ 0.54 & 6.46 $\pm$ 1.04 & ... & ... &  $-0.26~\pm~ 0.09$\\
52334 & 484 & 17.91 $\pm$ 1.12 & 13.34 $\pm$ 0.82 & 9.23 $\pm$ 0.50 & 6.02 $\pm$ 0.91 & ... & ...&  $-0.39~\pm~0.08$\\
52339 & 489  & ...& ...& 8.53 $\pm$ 0.57 &  ...& ...& ... & ...\\
52343 & 493 & 18.65 $\pm$ 0.95 & 11.48 $\pm$ 0.63 & 8.44 $\pm$ 0.45 & 4.73 $\pm$ 0.60 & 3.39 $\pm$ 0.42 & ... & $-0.53~\pm~0.08$\\ 
52351 & 501 & 16.44 $\pm$ 0.84& 9.03 $\pm$ 0.86 & 6.72 $\pm$ 0.43 & 4.22 $\pm$ 1.00 & 2.93 $\pm$ 0.57 & ...& $-0.53 ~\pm~ 0.04$\\  
52367 & 517 & 12.30 $\pm$ 0.68 & 7.31 $\pm$ 0.40 & 6.15 $\pm$ 0.34 & 4.66 $\pm$ 0.55 & 3.12 $ \pm$ 0.96 & ... & $-0.41 ~\pm~0.02$ \\
52385 & 535 & 9.30 $\pm$ 0.49 & 5.25 $\pm$ 0.41 & 3.66 $\pm$ 0.28 & ... &... &...& $-0.51 ~\pm~ 0.04$\\   
52393 & 543 & 7.91 $\pm$ 0.41 & 4.53 $\pm$ 0.30 & 2.88 $\pm$ 0.18 & 2.01 $\pm$ 0.35 & ... & ... & $-0.55~\pm~0.05$\\ 
52398 & 548 & 6.52 $\pm$ 0.35 & 3.69 $\pm$ 0.29 & ...& 1.70 $\pm $ 0.32 & ... & ... & $-0.51 ~\pm~ 0.06$\\
52412 & 562 & 7.66 $\pm$ 0.59 & 1.97 $\pm$ 0.13 & ... & ... & ...& ...& ...\\
52421 & 571 & 4.84 $\pm$ 0.31 & 1.95 $\pm$ 0.16 & ...& ...& ...& ...& ...\\   
52433 & 583 & 2.32 $\pm$ 0.13 & 1.43 $\pm$ 0.09 & ... & ...& ...& ... & ...\\
52443 & 593 & 2.79 $\pm$ 0.16 & 1.10 $\pm$ 0.13 & ...& ... & ...& ... & ...\\ 
52451 & 601 & 2.10 $\pm$ 0.11 & 1.40 $\pm$ 0.14 & 1.07 $\pm$ 0.26 & ... & ...& ...& $-0.35 ~\pm~ 0.02$ \\
52461 & 611 & 3.35 $\pm$ 0.21 & 1.01 $\pm$ 0.12 & 0.94 $\pm$ 0.20 & ...& ...& ...& $-0.87 ~\pm~ 0.15$\\ 
52475 & 625 & ...& ...& 0.68 $\pm$ 0.08 & ...& ...& ...& ...\\
52483 & 633 & 1.89 $\pm$ 0.11 & 1.02 $\pm$ 0.11 & 0.88 $\pm$ 0.14 & ...& ...& ...&  $-0.46 ~\pm~ 0.04$\\
52504 & 654 & 1.73 $\pm$ 0.15 & 0.99 $\pm$ 0.15 & 0.75 $\pm$ 0.10 &  ...& ...& ...& $-0.47 ~\pm~ 0.01$\\
52517 & 667 & 1.06 $\pm$ 0.08 & 0.61 $\pm$ 0.06 & 0.49 $\pm$ 0.05 & ... & ...& ...& $-0.44 ~\pm~ 0.01$ \\   
52519 & 669 & 2.09 $\pm$ 0.11 & 1.20 $\pm$ 0.15 & 0.92 $\pm$ 0.17 & ... & ...& ... & $-0.46 ~\pm~ 0.00$\\
52552 & 702 & ...& 0.78 $\pm$ 0.12 & 0.61 $\pm$ 0.11 & ...& ...& ... & ...\\
52575 & 725 & 1.46 $\pm$ 0.19 & 0.89 $\pm$ 0.07 & 0.49 $\pm$ 0.11 & ...& ...& ... & $-0.47 ~\pm~ 0.07$\\
52579 & 729 & ...& ...& 0.52 $\pm$ 0.12 &  ...& ...& 1.50 $\pm$ 0.31 & ...\\
52617 & 767 & ...& 1.15 $\pm$ 0.08 & 0.98 $\pm$ 0.14 & ...& ...& ...& ...\\ 
52629 & 779 &...& 0.80 $\pm$ 0.06 & 0.86 $\pm$ 0.12 & ...& ...& ...& ...\\
52637 & 787 & ...& 0.99 $\pm$ 0.11 & 0.88 $\pm$ 0.14 & ...& ...& ...& ...\\   
52642 & 792 & ...& 0.85 $\pm$ 0.11 & 0.71 $\pm$ 0.10 & ...& ...& ... & ...\\
52645 & 795 & ...&1.08 $\pm$ 0.12 & 0.84 $\pm$ 0.10 & ...& ...& ...& ...\\  
52647 & 797 &...& 0.77 $\pm$ 0.11 & 0.66 $\pm$ 0.16 & ...&...& ...& ...\\

\hline 
\end{tabular}
\begin{tablenotes}
\item `...' indicates no measurements for flux density for this epoch at that frequency 
\end{tablenotes}
\end{table*}

\begin{table*}
    \centering
    \contcaption{Flux densities and spectral indices of V445~Pup}
\begin{tabular}{lcccccccr}
\hline
$t$ & $t-t_0$ &\multicolumn{6}{c}{Radio flux densities (mJy)}& $\alpha$\\
 (MJD) & (Days) &1.43 GHz & 4.86 GHz & 8.46 GHz& 14.94 GHz & 22.46 GHz & 43.34 GHz&\\
\hline
52653 & 803 & ...& 0.84 $\pm$ 0.10 & 0.61 $\pm$ 0.10 & ...& ...& ...& ...\\ 
52659 & 809 & ...& 0.77 $\pm$ 0.07 & 0.65 $\pm$ 0.10 & ...& ...& ...& ...\\
52668 & 818 & ...& 1.22 $\pm$ 0.09 & 0.60 $\pm $ 0.07 & ...& ...& ...& ...\\
52676 & 826 & 2.85 $\pm$ 0.59 & 0.78 $\pm$ 0.09 & 0.77 $\pm$ 0.07 & ...& ...& ...& $-0.79 ~\pm~ 0.08$ \\
52693 & 843 & ...& 0.86 $\pm$ 0.11 & 0.63 $\pm$ 0.09 & ...& ...& ...& ...\\
52702 & 852 & ...& 0.90 $\pm$ 0.11 & 0.724 $\pm$ 0.13 & ... & ...& ...& ...\\
52709 & 859 &...&  0.93 $\pm$ 0.10 & 0.80 $\pm$ 0.12 & ...& ...& ...& ...\\
52711 & 861 & ...& 0.94 $\pm$ 0.09 & 0.74 $\pm$ 0.09 & ...&...&...& ...\\  
52717 & 867 &...& 0.84 $\pm$ 0.09 & 0.83 $\pm$ 0.10 & ...& ...& ...& ...\\
52724 & 874 & ...& 1.42 $\pm$ 0.18 & 1.15 $\pm$ 0.13 & ...& ...& ...& ...\\
52748 & 898 &...& 1.17 $\pm$ 0.11 & 0.66 $\pm$ 0.09 & ...& ...& ...& ...\\
52790& 940 & 2.21 $\pm$ 0.19 & 0.91 $\pm$ 0.09 & 0.24 $\pm$ 0.04& ...& ...& ...& $-0.93~\pm~ 0.47$\\
52795 & 945 & 2.53 $\pm$ 0.27 &0.78 $\pm$ 0.08 & ...& ...& ...&...& ...\\
52807 & 957 & 2.55 $\pm$ 0.24 & 0.90 $\pm$ 0.10 & ...& ...& ...& ...& ...\\
52821 & 971 & 2.49 $\pm$ 0.22 & 1.76 $\pm$ 0.22 & ...& ...& ...& ...& ...\\
52839 & 989 & ...& ...& 1.14 $\pm$ 0.10 & ...& ...& ...& ...\\
52842 & 992 & ...& 1.7 $\pm$ 0.15 & ...& ...& ...& ...&...\\ 
52846 & 996 & ...& 1.56 $\pm$ 0.15 & ...& ...& ...& ...&...\\  
52852 & 1002 & ...& ...& 0.65 $\pm$ 0.08 & ...& ...&...&..\\  
52863 & 1013 & 1.69 $\pm$ 0.16 & 1.37 $\pm$ 0.10 & 1.07 $\pm$ 0.08 & ...& ...& ...& $-0.25~\pm~ 0.08$\\  
52877 & 1027 & 2.22 $\pm$ 0.32 & 1.59 $\pm$ 0.23 & 1.49 $\pm$ 0.20 & ...& ...& ...& ...\\
52886 & 1036 & 2.67 $\pm$ 0.24 & 1.89 $\pm$ 0.31 & 1.01 $\pm$ 0.08 & ...& ...& ...& ...\\
52894 & 1044 & 1.49 $\pm$ 0.17 & 1.43 $\pm$ 0.16 & 0.87 $\pm$ 0.08 & ...& ...& ...& ...\\
52902 & 1052 & 1.02 $\pm$ 0.16 & ...& 1.10 $\pm$ 0.07 & ...& ...& ...& ...\\
52905 & 1055 & 1.49 $\pm$ 0.16 & 1.51 $\pm$ 0.16 & 1.00 $\pm$ 0.14 & ...& ...& ...& ...\\
52910 & 1060 & 2.05 $\pm$ 0.18 & 1.44 $\pm$ 0.12 & ...& ...& ...& ...& ...\\  
52912 & 1062 & ...&1.58 $\pm$ 0.20 & 0.87 $\pm$ 0.08 & ...& ...& ...& ...\\
52921 & 1071& ...& 1.09 $\pm$ 0.21 & ...&...& ...& ...&...\\
52934 & 1084 & ...& 1.38 $\pm$ 0.11 & 1.28 $\pm$ 0.13 & ...& ...& ...& ...\\
52946 & 1096 & ...& 1.32 $\pm$ 0.16 & 1.28 $\pm$ 0.17 & ...& ...& ...& ...\\
52953 & 1103 & ...& 1.34 $\pm$ 0.16 & 1.28 $\pm$ 0.12 & ...& ...& ...& ...\\
52962 & 1112 & ...& ...& 1.44 $\pm$ 0.15 & ...& ...& ...& ...\\
52967 & 1117 &...& 1.59 $\pm$ 0.15 & 1.30 $\pm$ 0.15 &...& ...& ...& ...\\
52988 & 1138 & 1.41 $\pm$ 0.25 & 1.82 $\pm$ 0.11& 1.46 $\pm$ 0.08 & ...& ...& ...& $-0.10 ~\pm~ 0.24$\\    
53001 & 1151 & 2.06 $\pm$ 0.10 & 1.72 $\pm$ 0.12 & 1.60 $\pm$ 0.15& ...& ...& ...& ...\\ 
53007 & 1157 & 3.06 $\pm$ 0.22 & 1.80 $\pm$ 0.18 & 1.71 $\pm$ 0.18 & ...& ...& ...& ...\\
53012 & 1162 & 2.01 $\pm$ 0.11 & 1.90 $\pm$ 0.20 & 1.62 $\pm$ 0.16 & ...& ...& ...& ...\\ 
53018 & 1168 & ... & 1.76 $\pm$ 0.13 & 1.56 $\pm$ 0.14 & ...& ...& ...& ...\\ 
53027 & 1177 &...& 1.74 $\pm$ 0.14 & 1.67 $\pm$ 0.15 & ...& ...& ...& ...\\    
53036 & 1186 & 2.17 $\pm$ 0.19 & 2.10 $\pm$ 0.21 & 1.50 $\pm$ 0.16 & ...& ...& ...& ...\\ 
53037 & 1187 & ...& ...& 1.54 $\pm$ 0.15 & 1.33 $\pm$ 0.38 & 1.05 $\pm$ 0.22 & 0.78 $\pm$ 0.18 & $-0.42~\pm~ 0.08$\\ 
53047 & 1197 &...& 2.09 $\pm$ 0.30 & 1.85 $\pm$ 0.19 & ...& 0.84 $\pm$ 0.17& 1.29 $\pm$ 0.37 & ... \\  
53059 & 1209 & ...& 1.53 $\pm$ 0.13 & 1.35 $\pm$ 0.12 & ...& 1.14 $\pm$ 0.27 & ...& $-0.20~ \pm~ 0.02$\\ 
53066 & 1216 & ...& 1.33 $\pm$ 0.12 & 1.30 $\pm$ 0.10 & ...& 0.84 $\pm$ 0.18 & ...& ...\\
53074 & 1224 & ...& 1.72 $\pm$ 0.10 & ...& ...& ...& ....& ...\\
53084 & 1234 & ...& 1.54 $\pm$ 0.17& 1.34 $\pm$ 0.13 & ...& ...& ...& ...\\
53086 & 1236 & ...& ...& 1.29 $\pm$ 0.15 & ...& ...& ...& ...\\
53096 & 1246 & ...& ...& 1.10 $\pm$ 0.12 & ...& ...& ...& ...\\ 
53102 & 1252 & ...& 1.15 $\pm$ 0.12 & 1.27 $\pm$ 0.12 &...& 0.73 $\pm$ 0.14 & ...& $-0.24~\pm~ 0.28$\\
53111 & 1261 & ...& 1.37 $\pm$ 0.20 & 1.04 $\pm$ 0.12 & ...& ...&...&...\\
53129 & 1279 & ...& 1.61 $\pm$ 0.20 & 1.14 $\pm$ 0.15 & ...& ...& ...& ...\\
53136 & 1286 & 2.27 $\pm$ 0.28 & 1.48 $\pm$ 0.10 & 1.18 $\pm$ 0.08 & ....& ...& ...&$-0.37 ~\pm~ 0.02$ \\   
53143 & 1293 & 0.94 $\pm$ 0.17 & 1.63 $\pm$ 0.20 & 1.13 $\pm$ 0.16 & ...& ...& ...& ...\\
53147 & 1297 & 3.45 $\pm$ 0.50 & ...& ...& ...& ...& ...& ...\\ 
53156 & 1306 & 1.49 $\pm$ 0.18 & 1.79 $\pm$ 0.22 & 1.52 $\pm$ 0.18 & ...& ...& ...& ...\\
53169 & 1319 & 2.51 $\pm$ 0.24 & 1.50 $\pm$ 0.16 & 1.23 $\pm$ 0.12 & ...& ...& ...& ...\\  
53182 & 1332 & ...& 1.01 $\pm$ 0.10 & 0.72 $\pm$ 0.08 & ...& ...& ...& ...\\
53188 & 1338 & ... & 1.29 $\pm$ 0.18 & 0.96 $\pm$ 0.13 & ...& ...& ...&...\\
53197 & 1347 & ...& ...& 0.89 $\pm$ 0.12 & 0.59 $\pm$ 0.11 &  ...& ...& ...\\
53204 & 1354 &...& 1.46 $\pm$ 0.12 & 1.17 $\pm$ 0.09 & 0.91 $\pm$ 0.19 & ...& ...&$-0.42~\pm~ 0.01$\\
53210 & 1360 & ...& 1.42 $\pm$ 0.18 & 1.25 $\pm$ 0.17 & ...& ...& ...& ...\\
53225 & 1375 & ...& 1.21 $\pm$ 0.16 & 1.07 $\pm$ 0.12 & 0.98 $\pm$ 0.21 & ...& ...&...\\ 

\hline 
\end{tabular}
\begin{tablenotes}
\item `...' indicates no measurements for flux density for this epoch at that frequency 
\end{tablenotes}
\end{table*}

\begin{table*}
\centering
 \contcaption{Flux densities and spectral indices of V445~Pup}
\begin{tabular}{lcccccccr}
\hline
$t$ & $t-t_0$ &\multicolumn{6}{c}{Radio flux densities (mJy)}& $\alpha$\\
 (MJD) & (Days) &1.43 GHz & 4.86 GHz & 8.46 GHz& 14.94 GHz & 22.46 GHz & 43.34 GHz&\\
\hline
53239 & 1389 & ...& 1.61 $\pm$ 0.18 & 1.37 $\pm$ 0.15 & 1.46 $\pm$ 0.28 & ...& ...& ...\\
53258 & 1408 & ...& 1.26 $\pm$ 0.19 & 1.45 $\pm$ 0.31 & 1.16 $\pm$ 0.28 & ...& ...& $-0.03 ~\pm~ 0.17$\\  
53272 & 1422 & ...& 1.07 $\pm$ 0.11 &1.43 $\pm$ 0.19 &...& ...& ...& ...\\
53280 & 1430 & ...& ...& 1.51 $\pm$ 0.19 & ...& ...&...& ...\\
53287 & 1437 &...& ...& 1.56 $\pm$ 0.14 &  ...& ...& ...& ...\\
53295 & 1445 &...&...& 1.71 $\pm$ 0.15 & ...&...&...& ...\\
53310 & 1460 & ...& 1.61 $\pm$ 0.14 & 1.09 $\pm$ 0.16 & ...& ...& ...& ...\\
53312 & 1462 &... & 1.12 $\pm$ 0.12 & 1.38 $\pm$ 0.14 &...&...&...&...\\
53317 & 1467 & ...&...& 1.39 $\pm$ 0.13 & ...&...&...&...\\
53321 & 1471 & ...&...&1.54 $\pm$ 0.13 &...&...&...&...\\
53329 & 1479 & ...&...&1.21 $\pm$ 0.15 &...&...&...&...\\
53338 & 1488 &...&...& 1.26 $\pm$ 0.17 & ...&...&...&...\\
53361 & 1511 & ...&...&1.27 $\pm$ 0.19& ...&...&...&...\\
53370 & 1520 & ...& ...& 1.46 $\pm$ 0.19 & ...&...&...& ...\\
53376 & 1526 & ...&...& 1.27 $\pm$ 0.11 & ...&...&...&..\\
53386 & 1536 & ...&...&1.21 $\pm$ 0.17& ...&...&...&...\\
53394 & 1544 & ...&...&1.38 $\pm$ 0.15 & ...&...&...&...\\
53399 & 1549 &...& 1.36 $\pm$ 0.15 & 1.23 $\pm$ 0.14 & ...&...&...&...\\
53404 & 1554 &...& ...& 1.41 $\pm$ 0.16 & ...& ...&...& ...\\
53427 & 1577 &...& 1.28 $\pm$ 0.14 & 1.20 $\pm$ 0.13 & ...&...&...&...\\
53442 & 1592 & 1.15 $\pm$ 0.12 & 0.94 $\pm$ 0.07 & 0.87 $\pm$ 0.08 & ...& ...& ...& $-0.16~\pm~ 0.01$\\  
53449 & 1599 & 0.89 $\pm$ 0.08 & 1.05 $\pm$ 0.08 & 1.06 $\pm$ 0.12 & ...&...&...& ...\\ 
53454 & 1604 & 1.10 $\pm$ 0.11 & 1.32 $\pm$ 0.18 & 1.02 $\pm$ 0.16 & ...&...&...& ...\\ 
53464 & 1614 & ...& 1.17 $\pm$ 0.09 & ...& ...&...&...& ...\\
53475 & 1625 & ...&0.96 $\pm$ 0.12 & 0.86 $\pm$ 0.12 & ...&...&...& ...\\
53486 & 1636 &...& 1.21 $\pm$ 0.15 & 0.93 $\pm$ 0.11 & ...&...&...&...\\
53497 & 1647 & ...& 1.13 $\pm$ 0.14 & 0.72 $\pm$ 0.10 & ...&...&...&...\\
53504 & 1654 &...& 0.90 $\pm$ 0.16 & 0.80 $\pm$ 0.12 & ...&...&...&...\\
53520 & 1670 & ...& 0.94 $\pm$ 0.12 & ...&...&...&...&...\\
53525 & 1675 & ...& 0.99 $\pm$ 0.12 & ...&...&...&...&...\\
53534 & 1684 & ...& ...& 0.62 $\pm$ 0.074 &...&...&...&...\\
53540 & 1690  & ...& ...& 0.83 $\pm$ 0.09 & ...&...&...&...\\
53545 & 1695 & ...& 1.01 $\pm$ 0.13& 0.76 $\pm$ 0.13 & ...&...&...&...\\
53554 & 1704 & 1.42 $\pm$ 0.10 & 0.73 $\pm$ 0.07 & 0.81 $\pm$ 0.11 & ...&...&...& ...\\   
53561 & 1711 & ...& ...& 0.64 $\pm$ 0.08 &...&...&...&...\\
53566 & 1716 & ...& 0.77 $\pm$ 0.08 & ...& ...&...&...&...\\ 
53575 & 1725 &...& 0.68 $\pm$ 0.08 & 0.51 $\pm$ 0.06 & ...& ...&...& ...\\
53585 & 1735 &...& ...& 0.47 $\pm$ 0.08 & ...& ...&...& ...\\
53589 & 1739 & ...&...& 0.78 $\pm$ 0.09 &...& ...& ...& ...\\
53610 & 1760 & ...& 0.86 $\pm$ 0.09 & 0.53 $\pm$ 0.07 & ...&...&...&...\\
53619 & 1769 & ...& 1.12 $\pm$ 0.14 &  0.53 $\pm$ 0.05 & ...&...&...& ...\\
53639 & 1789 & ...& 0.73 $\pm$ 0.08 & ...&...&...&...&...\\
53641 & 1791 & ...& 0.58 $\pm$ 0.05 & 0.68 $\pm$ 0.04 & ...&...&...& ...\\
53645 & 1795 &...& 0.65 $\pm$ 0.05 & 0.54 $\pm$ 0.05 & ...&...&...&...\\
53649 & 1799 &...&...& 0.61 $\pm$ 0.09 & ...&...&...&...\\
53660 & 1810 & ...&...& 0.75 $\pm$ 0.07 & ...&1.10 $\pm$ 0.30 &...&...\\
53666 & 1816 &...&...& 0.66 $\pm$ 0.06 & ...&...&...&...\\
53675 & 1825 & ...&...& 0.64 $\pm$ 0.07& ...&...&...&...\\
53683 & 1833 & ...&0.62 $\pm$ 0.11 &0.58 $\pm$ 0.09 & ...&...&...&...\\
53689 & 1840& ...& ...& 0.57 $\pm$ 0.08& ...&...&...&...\\
53696 & 1846 &...&...& 0.54 $\pm$ 0.07 & ...&...&...&...\\
53703 & 1853 & ...&...& 0.74 $\pm$ 0.09 & ...& ...&...& ...\\
53705 & 1855 & ...&0.51 $\pm$ 0.05 & 0.55 $\pm$ 0.05 & ...&...&...&...\\      
53715 & 1865 & ...& 0.82 $\pm$ 0.18 & 0.64 $\pm$ 0.08 & ...&...&...&...\\
53724 & 1874 & ...& 0.60 $\pm$ 0.12 & 0.53 $\pm$ 0.08 & ...&...&...&...\\
53748 & 1898 &...&...& 0.59 $\pm$ 0.08 & ...&...&...&...\\
53750 & 1900 & ...& ...& 0.48 $\pm$ 0.08 & ...& ...&...&...\\
53757 & 1907 & ...& 0.47 $\pm$ 0.09 & ...& ...&...&...&...\\
53775 & 1925 & ...& ...& 0.43 $\pm$ 0.09 & ...&...&...& ...\\
53809 & 1959 & 0.54 $\pm$0.18 &...& 0.40 $\pm$ 0.09 & ...&...&...&...\\
53816 & 1966 & 0.51 $\pm$0.12 &...& ... & ...&...&...&...\\
53832 & 1982 & 0.61 $\pm$0.22 &...& ... & ...&...&...&...\\
53863 & 2013 & ...&...& 0.66 $\pm$ 0.09 & ...&...&...&...\\

\hline 
\end{tabular}
\begin{tablenotes}
\item `...' indicates no measurements for flux density for this epoch at that frequency 
\end{tablenotes}
\end{table*}

\begin{table*}
\centering
 \contcaption{Flux densities and spectral indices of V445~Pup}
\begin{tabular}{lcccccccr}
\hline
$t$ & $t-t_0$ &\multicolumn{6}{c}{Radio flux densities (mJy)}& $\alpha$\\
 (MJD) & (Days) &1.43 GHz & 4.86 GHz & 8.46 GHz& 14.94 GHz & 22.46 GHz & 43.34 GHz&\\
\hline
53865 & 2015 & ...&...&0.37 $\pm$ 0.05 & ...&...&...&...\\
53953 & 2103 & ...& ...& 0.21 $\pm$ 0.03 & ...&....&...&...\\
53957 & 2107 & ...&0.42 $\pm$ 0.08 & ...& ...&...&...&...\\
53979 & 2129 & ...& 0.30 $\pm$ 0.06 & ...&...&...&...& ...\\
54123 & 2273 &...&...& 0.29 $\pm$ 0.06& ...&...&...&...\\
54374 & 2524 & ...&...& 0.22 $\pm$ 0.06 & ...&...&...&...\\
54482 & 2632 &...& ...& 0.26 $\pm$ 0.09 & ...&...&...&...\\
54554 & 2704 &...&...& 0.19 $\pm$ 0.04& ...&...&...&...\\  

\hline 
\end{tabular}
\begin{tablenotes}
\item `...' indicates no measurements for flux density for this epoch at that frequency 
\end{tablenotes}
\end{table*}

\bsp	
\label{lastpage}
\end{document}